\documentclass[acmtosem,screen]{acmart}

\usepackage[utf8]{inputenc}
\usepackage[english]{babel}

\usepackage{graphicx}
\DeclareGraphicsExtensions{.pdf,.png,.jpg}

\usepackage{booktabs}
\usepackage{multirow}
\usepackage{enumitem}
\usepackage{makecell}
\usepackage{relsize}
\usepackage{caption}
\usepackage{tabularx}
\usepackage{array}
\usepackage{pifont}
\usepackage{longtable}
\usepackage{pdflscape}
\usepackage{xcolor}
\usepackage{ragged2e}
\usepackage{colortbl}
\usepackage{microtype}
\usepackage{xltabular}
\usepackage{adjustbox}
\usepackage{cprotect}
\usepackage{seqsplit}
\usepackage{fvextra}
\usepackage[T1]{fontenc}

\usepackage[dvipsnames]{xcolor}
\usepackage[most]{tcolorbox}
\usepackage[normalem]{ulem}
\usepackage{xcolor,colortbl}

\definecolor{darkgreen}{RGB}{0,100,0}
\definecolor{codegreen}{rgb}{0,0.5,0}
\definecolor{codepurple}{rgb}{0.58,0,0.82}
\definecolor{codegray}{rgb}{0.5,0.5,0.5}
\usepackage{listings}
\lstdefinestyle{mystyle}{
  commentstyle=\color{codegreen},
  keywordstyle=\bfseries,
  stringstyle=\color{codepurple},
  basicstyle=\ttfamily\scriptsize,
  breaklines=true,
  captionpos=b,
  keepspaces=true,
  tabsize=2
}
\usepackage{algorithm}
\usepackage{algpseudocode}

\usepackage{enumitem}

\usepackage{tikz}
\usetikzlibrary{arrows.meta,positioning,shapes,calc} 

\definecolor{codebg}{gray}{0.97}
\definecolor{codecomment}{gray}{0.40}
\definecolor{codekw}{rgb}{0.0,0.0,0.55}
\lstdefinestyle{kernelC}{
  language=C,
  basicstyle=\ttfamily\small,
  commentstyle=\color{codecomment}\itshape,
  keywordstyle=\color{codekw}\bfseries,
  backgroundcolor=\color{codebg},
  numbers=none,
  xleftmargin=4pt,
  xrightmargin=4pt,
  frame=single,
  framerule=0.4pt,
  rulecolor=\color{black!30},
  breaklines=true,
  breakatwhitespace=false,
  columns=flexible,
  keepspaces=true,
  showstringspaces=false,
  tabsize=4,
  aboveskip=6pt,
  belowskip=6pt,
  morekeywords={SEC,bpf_ringbuf_reserve,bpf_ringbuf_submit,
    __xdp_return,skb_frag_page,skb_frag_size,page_address,
    xsk_buff_get_tail,xsk_buff_del_tail,update_effective_progs,
    list_del,bpf_link_free,purge_effective_progs,
    MEM_TYPE_XSK_BUFF_POOL,PTR_OR_NULL,bpf_xdp_adjust_tail,
    bpf_xdp_shrink_data},
  literate={→}{$\rightarrow$}1 {←}{$\leftarrow$}1
           {↑}{$\uparrow$}1 {↔}{$\leftrightarrow$}1,
}

\lstdefinestyle{diagram}{
  basicstyle=\ttfamily\small,
  backgroundcolor=\color{codebg},
  numbers=none,
  xleftmargin=4pt,
  xrightmargin=4pt,
  frame=single,
  framerule=0.4pt,
  rulecolor=\color{black!30},
  breaklines=true,
  columns=flexible,
  keepspaces=true,
  aboveskip=6pt,
  belowskip=6pt,
  literate={→}{$\rightarrow$}1 {←}{$\leftarrow$}1
           {↑}{$\uparrow$}1,
}

\newboolean{COMMENTSON} 
\setboolean{COMMENTSON}{true}   
\ifthenelse{\boolean{COMMENTSON}}
{

}

\definecolor{DarkOrange}{rgb}{0.8,0.3,0.0} 
\definecolor{DarkCyan}{rgb}{0.0, 0.55, 0.55}
\definecolor{codegreen}{rgb}{0,0.6,0}
\definecolor{codegray}{rgb}{0.5,0.5,0.5}
\definecolor{codepurple}{rgb}{0.58,0,0.82}
\definecolor{backcolour}{rgb}{0.95,0.95,0.92}

\newcommand{\papertitle}{eBPF Security in the Wild: Structural Concentration, Failure Mechanisms, and Discovery Gaps}
\newcommand{\papertitleshort}{eBPF Security in the Wild: Structural Concentration, Failure Mechanisms, and Discovery Gaps}

\newcommand{\paperkeywords}{eBPF, eBPF security, eBPF vulnerabilities}

\newcommand{\authorAname}{Baihong Chen}
\newcommand{\authorAaffil}{Utah State University}
\newcommand{\authorAemail}{b.chen@usu.edu}

\newcommand{\authorBname}{Wen Li}
\newcommand{\authorBaffil}{Utah State University}
\newcommand{\authorBemail}{awen.li@usu.edu}

\newcommand{\authorCname}{Hua Ming}
\newcommand{\authorCaffil}{University of Michigan}
\newcommand{\authorCemail}{huaming@umich.edu}

\newcommand{\authorDname}{Weifeng Pan}
\newcommand{\authorDaffil}{Zhejiang Gongshang University}
\newcommand{\authorDemail}{wfpan@zjgsu.edu.cn}

\newcommand{\authorEname}{Tian Xie}
\newcommand{\authorEaffil}{Utah State University}
\newcommand{\authorEemail}{tian.xie@usu.edu}

\newcommand{\authorFname}{Xiaojun Qi}
\newcommand{\authorFaffil}{Utah State University}
\newcommand{\authorFemail}{xiaojun.qi@usu.edu}

\renewcommand{\arraystretch}{1.3}

\newcommand{\cmark}{\ding{51}}
\newcommand{\pmark}{\ding{119}}
\newcommand{\xmark}{\ding{55}}

\newcommand{\blackcircleone}[1]{%
  \tikz[baseline=(char.base)]{
    \node[shape=circle, fill=black, text=white, inner sep=1pt] (char) {\small #1};
  }%
}

\setcopyright{acmcopyright}

\begin{document}

\title[\papertitleshort]{\papertitle}

\author{\authorAname}
\affiliation{%
  \institution{\authorAaffil}
  \country{USA}
}
\email{\authorAemail}
\authornote{First author}

\author{\authorCname}
\affiliation{%
  \institution{\authorCaffil}
  \country{USA}
}
\email{\authorCemail}

\author{\authorDname}
\affiliation{%
  \institution{\authorDaffil}
  \country{China}
}
\email{\authorDemail}

\author{\authorEname}
\affiliation{%
  \institution{\authorEaffil}
  \country{USA}
}
\email{\authorEemail}

\author{\authorFname}
\affiliation{%
  \institution{\authorFaffil}
  \country{USA}
}
\email{\authorFemail}

\author{\authorBname}
\affiliation{%
  \institution{\authorBaffil}
  \country{USA}
}
\email{\authorBemail}
\authornote{Corresponding author}


\begin{abstract}

Extended Berkeley Packet Filter (eBPF) is a security-critical in-kernel execution framework, yet its vulnerability landscape remains fragmented across components, semantic gaps, and testing techniques. We present an empirical study of observed eBPF vulnerabilities. We construct a multi-source dataset from Linux kernel fixing commits, syzbot reports, and public CVE/NVD records, and analyze it through a unified framework covering structural concentration, mechanism-level failure modes, architectural distribution, and discovery gaps in representative techniques.

Our results show that the observed eBPF vulnerability landscape is structurally concentrated rather than broadly dispersed across many unrelated weakness types. The dominant portion is associated with a limited set of recurring system-level failures, especially in runtime execution, concurrency, object lifecycle management, and semantic inconsistencies across trusted stages. These failures are unevenly distributed across the eBPF pipeline: Runtime is the dominant exposure surface, whereas the Verifier and JIT are lower-frequency but structurally distinct security boundaries. A rubric-based comparison of representative techniques and a version-aligned Syzkaller case study on Linux v5.10 show that, despite visible raw coverage of Runtime, Verifier, and JIT, effective exploration is semantically narrow, and observed discoveries concentrate in a small subset of Runtime failures. Overall, raw coverage alone provides an incomplete view of discovery effectiveness.

\end{abstract}

\maketitle

\section{Introduction}

Extended Berkeley Packet Filter (eBPF) has emerged as a powerful in-kernel execution framework that enables flexible programmability for networking, observability, and security enforcement in modern operating systems.
By allowing user-defined programs to execute inside the kernel while preserving key safety and performance properties, eBPF has become an increasingly important part of the Linux software stack~\cite{ebpfdocs,linuxkernel}.
At the same time, this programmability introduces new security challenges.
The eBPF execution pipeline spans multiple tightly coupled components, including the verifier, the just-in-time (JIT) compiler, and the Runtime subsystem.
Each of these components is responsible for enforcing different correctness and safety properties, and vulnerabilities may arise not only from flaws within one component, but also from semantic mismatches across stages, incomplete enforcement of assumptions, or unsafe interactions between mechanisms.
As eBPF continues to expand in functionality and deployment scope, understanding the structure of its real-world vulnerability space becomes increasingly important.

Prior work has examined eBPF security from several complementary directions.
A first line of research studies component-specific security properties, especially in the verifier, including correctness, soundness, and enforcement limitations~\cite{gershuni2019simple,bhat2022formal,sun2024validating}.
A second line of work focuses on cross-stage semantic inconsistencies, particularly mismatches between verifier assumptions and downstream execution in the JIT or Runtime~\cite{nelson2020specification,jia2025rex,peng2024toss}.
A third line of work explores dynamic testing and fuzzing for eBPF.
General-purpose kernel fuzzing infrastructures such as \textit{syzkaller}~\cite{syzkaller} and \textit{syzbot}~\cite{syzbot} have played an important practical role in exposing Linux kernel defects, including eBPF-related ones.
More specialized systems have targeted particular stages or mechanisms.
For example, \textit{Buzzer}~\cite{google_buzzer} focuses on verifier-oriented exploration, while \textit{BRF}~\cite{hung2024brf} is designed to improve verifier pass rates and increase reachability into the eBPF Runtime.
Together, these studies provide valuable insights into eBPF security, but most focus on one component, one semantic gap, or one testing technique at a time.
As a result, 
the current literature provides only a fragmented view of eBPF security.
It remains unclear how real-world eBPF vulnerabilities 
are distributed across architectural components, 
what dominant system-level failures underlie them, 
and how well the current discovery techniques align 
with that real-world vulnerability space.
Answering these questions requires moving beyond isolated mechanism studies 
toward a unified empirical perspective on real-world vulnerabilities.

In this paper, we present a systematic empirical study of observed eBPF vulnerabilities.
Rather than evaluating individual Verifier, JIT, or Runtime mechanisms in isolation, 
we study the eBPF vulnerability landscape as an empirical structure.
To do so, 
we construct a dataset grounded in Linux kernel fixing commits, syzbot reports, and public vulnerability records, 
and analyze these cases along both mechanism-level and architectural dimensions.
This allows us to examine how vulnerabilities concentrate in practice, 
what recurring failure mechanisms drive that concentration, 
how those mechanisms are distributed across the eBPF architecture, 
and how well representative current discovery techniques align with that observed structure.
Our analysis reveals a coherent empirical picture of the observed eBPF vulnerability landscape captured by our reconstructed dataset. 
The main findings of the study are as follows:

\begin{itemize}[leftmargin=*]
\item \textbf{Clear structural concentration.}
Rather than being uniformly distributed, 
the observed eBPF vulnerabilities in our reconstructed dataset concentrate in a relatively small number of dominant weakness categories.

\item \textbf{Recurring mechanism-level failures in the dominant sampled portion.}
The dominant sampled portion of this concentration is associated with a limited set of recurring system-level failure mechanisms, 
rather than a broad collection of unrelated defect types.

\item \textbf{Non-uniform architectural distribution.}
These dominant observed failures are unevenly distributed across the eBPF pipeline. 
In our dataset, Runtime-related failures account for the largest observed portion, 
while Verifier-, JIT-, and cross-component issues represent smaller but structurally distinct parts of the landscape.

\item \textbf{Uneven support among representative current techniques.}
When viewed against this empirical structure, 
the representative discovery techniques examined in this study show uneven support across dominant failure categories. 
Our rubric-based design-level comparison suggests stronger support for some Runtime-related categories than for several Verifier-, JIT-, and cross-component categories, 
while the version-aligned Syzkaller case study shows that, in the evaluated Linux v5.10 setting, 
visible raw coverage does not necessarily imply broad semantic exploration, 
and observed discoveries are concentrated in a limited subset of Runtime-related failures.
\end{itemize}

\vspace{2pt}
\noindent
Taken together, 
these findings suggest that the observed eBPF vulnerability landscape is not only concentrated, 
but also structured in ways that current discovery and evaluation practices do not fully capture. 
This leads to several broader implications for eBPF security research:

\begin{itemize}[leftmargin=*]
\item \textbf{eBPF weaknesses are best understood as a structured system-level problem.}
Because the observed landscape is shaped by recurring failure patterns and non-uniform architectural concentration, 
it is better understood as a structured system-level problem 
than as a collection of evenly distributed implementation mistakes.

\item \textbf{Raw coverage alone is not a sufficient measure of practical discovery effectiveness.}
Because several structurally important vulnerability categories in the observed landscape remain weakly explored in practice despite visible code reach, 
raw coverage alone can provide an incomplete picture of practical discovery capability~\cite{291011}.

\item \textbf{Future techniques may need stronger mechanism- and architecture-aware support.}
Since the dominant observed failures are concentrated in particular mechanisms and architectural regions, 
more effective eBPF security analysis will likely require techniques 
that are better aligned with those dominant failure mechanisms, 
more aware of architectural concentration, 
and more capable of exposing vulnerabilities beyond the currently overrepresented subset of Runtime-related failures.
\end{itemize}

\section{Background and Related Work}

This section introduces the eBPF execution pipeline and security model 
needed to understand our study, 
and then situates this work relative to prior research 
on eBPF security analysis, fuzzing, and empirical vulnerability studies.

\subsection{Background: eBPF Execution Flow and Trust Model}

eBPF extends the Linux kernel with programmable logic 
that is loaded from user space and executed inside the kernel.
Because eBPF programs cross the user/kernel boundary and can interact with sensitive kernel state, 
their security depends not on a single defense point, 
but on a staged pipeline of validation, translation, and execution~\cite{ebpfdocs,zhong2025depsurf}.

\begin{figure}[!ht]
\centering
\includegraphics[width=0.9\linewidth]{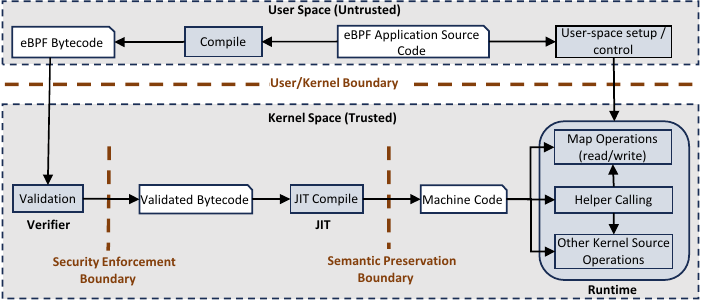}
\caption{eBPF execution workflow and trust model.}
\label{fig:work_flow_and_trust_model}
\end{figure}

Figure~\ref{fig:work_flow_and_trust_model} illustrates 
this workflow and the corresponding trust model.
An eBPF program is first written in a high-level language 
and compiled into eBPF bytecode in user space.
This bytecode is then submitted to the kernel, 
where it is checked by the \textbf{Verifier}.
If accepted, 
the program is translated by the \textbf{JIT} compiler into machine code.
The resulting code finally executes in the \textbf{Runtime}, 
where it interacts with maps, helper functions, and other kernel subsystems.
These stages play distinct security roles.

\noindent
\textbf{Verifier.}
The Verifier is the primary security-enforcement boundary.
Its task is to reject untrusted programs that may violate kernel safety requirements, 
such as invalid memory access, unsound pointer usage, illegal helper invocation, or unsafe control flow.
Because the Verifier reasons over abstract program states rather than concrete executions, 
its correctness depends on whether its semantic models, type rules, and state propagation logic remain sound and complete.

\noindent
\textbf{JIT.}
The JIT is the semantic-preservation boundary between verified bytecode and machine execution.
Even if the Verifier approves a program, security still depends on whether the architecture-specific JIT translation
faithfully preserves the semantics that were validated at the bytecode level.
JIT defects can therefore create a gap between verifier-approved behavior and actual machine-level execution.

\noindent
\textbf{Runtime.}
The Runtime is the primary execution exposure surface.
At this stage, eBPF programs interact with concrete kernel objects, helper functions, maps, synchronization primitives, 
and surrounding subsystems.
As a result, many practical security risks arise not from abstract reasoning failures, 
but from execution-time behaviors such as memory misuse, lifecycle inconsistencies, concurrency errors, 
or unsafe cross-subsystem interactions.

\subsection{Related Work}

Prior work on eBPF security has largely focused on particular components, 
specific semantic gaps, or individual testing techniques, 
rather than systematically characterizing the real-world vulnerability landscape 
across the full eBPF pipeline. We organize the most relevant prior work into three categories.

\subsubsection{Component-Specific Security Analysis}

A substantial body of prior work has studied individual eBPF components, 
either by analyzing their security properties directly or 
by developing testing techniques targeted at specific pipeline stages.
Verifier-centered studies examine 
whether the Verifier soundly enforces safety constraints over untrusted programs, 
including the precision of abstract interpretation, pointer reasoning, and state modeling. 
Prior analyses and validation efforts have shown that 
subtle inconsistencies in abstract state tracking can cause the Verifier to accept unsafe programs~\cite{gershuni2019simple,bhat2022formal,sun2024validating,sun2024ebpfverifier,lyu2025ebpfmisbehavior}. 
These works establish the Verifier as a critical security boundary, 
but focus primarily on reasoning soundness and verifier-specific defects.

Another line of work studies the correctness of eBPF JIT compilers. 
Because JIT backends translate verified bytecode into architecture-specific machine code, 
their correctness depends on faithfully 
preserving the semantics already approved by the Verifier. 
Prior formal-methods work has shown that JIT implementations 
can contain subtle architecture-dependent translation bugs and 
has proposed specification and verification techniques 
for reasoning about JIT correctness~\cite{nelson2020specification}. 
These studies highlight the JIT as an important boundary for semantic preservation, 
but remain focused on translation correctness 
rather than on the broader distribution of JIT-related vulnerabilities.

Recent work has also examined eBPF security through dynamic testing and fuzzing. 
General-purpose kernel fuzzing infrastructures 
such as syzkaller and syzbot have played an important practical role in exposing Linux kernel defects, including eBPF-related ones~\cite{syzbot,syzkaller}. 
More specialized eBPF-oriented systems have targeted particular mechanisms or stages~\cite{mohamed2023ebpfsecurity,spinner}. 
For example, Buzzer focuses on verifier-oriented exploration, 
while BRF is designed to improve verifier pass rates and 
increase reachability into the eBPF Runtime~\cite{hung2024brf}. 
These studies demonstrate the value of dynamic testing for exposing eBPF-related defects, 
but they primarily emphasize testing capability and bug-finding effectiveness 
rather than the broader structure of real-world vulnerability distributions.

Collectively, 
these component-specific studies provide important insights 
into individual security boundaries of the eBPF pipeline, 
but they do not provide a unified empirical view across components.

\subsubsection{Cross-Component and Semantic-Gap Studies}

Beyond single-component analysis, 
some recent work has highlighted semantic gaps and 
cross-stage inconsistencies across the eBPF stack. 
Prior research has shown that mismatches may arise between higher-level safety expectations, 
verifier models, and actual implementation behavior across different stages of the system~\cite{jia2025rex,peng2024toss}. 
This perspective is important because some eBPF vulnerabilities 
do not belong cleanly to a single component; 
instead, 
they emerge when guarantees assumed at one stage are not preserved 
by another stage or by surrounding subsystem behavior.
These studies are especially relevant to our work 
because they motivate viewing eBPF security as a pipeline problem 
rather than as a collection of isolated verifier, JIT, or runtime bugs. 
However, 
existing work in this direction still tends 
to focus on specific semantic inconsistencies or particular classes of cross-component defects, rather than systematically relating such failures 
to the broader real-world vulnerability landscape.

\subsubsection{Empirical and Broader Security Perspectives}

Broader systematization and survey work has framed eBPF 
as a security-critical extension mechanism 
whose risks span memory safety, isolation, verification, and execution-time behavior~\cite{huang2025sok}. 
These works clarify the overall challenge space and 
motivate the need for systematic security analysis, 
but they do not provide an evidence-guided empirical study of 
how real-world eBPF vulnerabilities 
are distributed across weakness classes, mechanisms, and architectural stages.
More generally, 
empirical software-security studies often characterize vulnerability populations 
through taxonomies, root-cause analysis, 
and distributional measurements. 
Our work brings that empirical perspective to eBPF, 
where prior studies have more often focused on component correctness, 
semantic inconsistencies, or testing techniques in isolation.

\subsection{Positioning of This Study}

Prior work has provided important insights into 
component-specific security properties, cross-stage semantic inconsistencies, 
and dynamic testing techniques for eBPF. 
However, 
most existing studies focus on one component, one semantic gap, 
or one testing technique at a time. 
By contrast, 
this study takes a unified empirical perspective on real-world eBPF vulnerabilities. 
Rather than asking only whether a specific verifier, JIT, 
or runtime mechanism is secure, 
we ask how real-world vulnerabilities are structurally distributed, 
what dominant system-level failures underlie them, 
how those failures are situated across the eBPF architecture, 
and where representative current discovery techniques remain insufficient in practice.
\section{Study Goals and Overview}\label{sec:goal}

This study conducts a systematic empirical analysis of real-world eBPF vulnerabilities 
based on a unified dataset constructed through evidence-guided attribution, cross-source reconstruction, 
and validated classification procedures.
Rather than treating observed vulnerabilities as isolated cases, we aim to understand how they are organized, 
what recurring failures they reflect, 
how these failures are situated within the eBPF architecture, 
and where current discovery practices remain insufficient.
Accordingly, 
the study is organized around four connected analytical dimensions:
structural concentration,
mechanism-level failure modes,
architectural concentration across major components and execution stages,
and discovery gaps in representative current techniques.
By connecting these dimensions within a unified empirical framework,
the study links observed vulnerability patterns to their underlying causes,
architectural distribution,
and implications for current vulnerability discovery practice.
To systematically structure this investigation, we formulate four research questions as follows:

\begin{itemize}
  \item \textbf{RQ1: What structural concentration patterns characterize real-world eBPF vulnerabilities?}
  This question focuses on the overall structural organization of real-world eBPF vulnerabilities.
  Its goal is to characterize the vulnerability landscape empirically and 
  to determine which observed patterns warrant deeper mechanism-level analysis in the later parts of the study.

  \item \textbf{RQ2: What mechanism-level failure modes underlie the dominant real-world eBPF vulnerabilities?}
  Building on the structural concentration, 
  this question focuses on the underlying system failures associated 
  with the dominant vulnerability patterns. 
  Its goal is to establish a mechanism-level understanding of the recurring failures reflected 
  in the dominant portion of the observed eBPF vulnerability landscape and 
  to explain the major patterns identified above.

  \item \textbf{RQ3: How are the dominant failure mechanisms distributed across major eBPF components and execution stages?}
  Building on the mechanism-level understanding,
  this question focuses on how the dominant failure mechanisms are situated across the eBPF architecture.
  Its goal is to characterize how recurring failures concentrate across major components 
  and execution stages in the eBPF security model.

  \item \textbf{RQ4: To what extent do representative existing techniques cover the dominant real-world eBPF vulnerability patterns?}
  Building on the findings of RQ2 and RQ3, 
  this question examines how well existing representative techniques 
  align with the major vulnerability patterns observed in our reconstructed dataset.
  Its goal is to summarize the capability boundaries of representative current techniques and 
  identify important discovery gaps relative to the dominant patterns characterized in this study, 
  with \textit{Syzkaller}~\cite{syzkaller} used as a case study for in-depth empirical analysis.
\end{itemize}

Collectively,
these four questions provide a unified framework for analyzing real-world eBPF vulnerabilities, 
from their overall distribution and underlying failure mechanisms 
to their architectural concentration and the limitations of current discovery techniques.
The questions are ordered from landscape characterization to mechanism explanation, architectural interpretation, and discovery-gap assessment.

\section{Methodology}\label{sec:method}

This section presents the empirical design of the study.
We first introduce the overall workflow used to construct and analyze the dataset,
and then describe each stage in detail.

\subsection{Overview of Study Methodology}
\label{sec:method:overview}

\begin{figure}[!ht]
\centering
\includegraphics[width=1\linewidth]{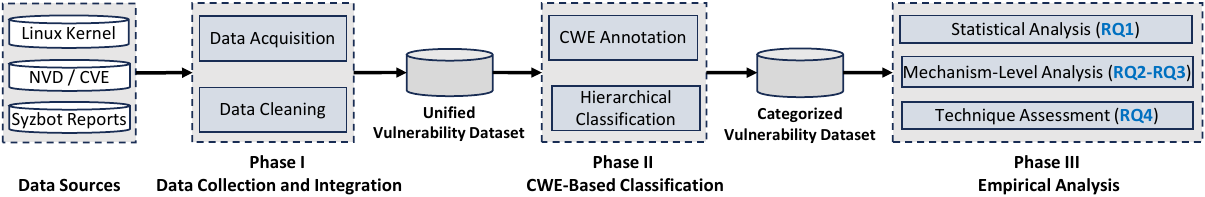}
\caption{Overall workflow of the study. Multi-source records from the Linux kernel, NVD/CVE, and syzbot are integrated into a unified vulnerability dataset, organized through CWE-based classification into a categorized vulnerability dataset, and then analyzed through statistical analysis, mechanism-level analysis, and technique assessment.}
\label{fig:overview}
\end{figure}

Figure~\ref{fig:overview} shows the overall workflow of the study.
The workflow proceeds through three connected stages:
Phase I integrates Linux kernel~\cite{linuxkernel}, syzbot~\cite{syzbot}, and NVD/CVE~\cite{nvd} records into a unified vulnerability dataset. 
Phase II assigns CWE labels and organizes the dataset under the CWE-1000 hierarchy~\cite{cwe1000}. 
Phase III uses the categorized dataset for structural analysis (RQ1), 
mechanism and architectural analysis (RQ2–RQ3), 
and technique assessment (RQ4).
Overall, the workflow forms a single empirical pipeline connecting multi-source data construction,
hierarchical vulnerability classification,
and the empirical analyses presented in the remainder of the paper.

\subsection{Data Sources} \label{sec:method:datasource}

This subsection defines the empirical evidence base of the study.
To support a systematic analysis of real-world eBPF vulnerabilities, 
we construct the dataset by integrating three complementary data sources:
vulnerability-fixing commits from the Linux kernel mainline repository~\cite{linuxkernel},
fixed reports from syzbot~\cite{syzkaller,syzbot},
and CVE records from the National Vulnerability Database (NVD)~\cite{nvd}.
Together, these sources capture three distinct but connected aspects of vulnerability evidence:
code-level remediation~\cite{li2017securitypatches,liu2025disclosure}, automated discovery~\cite{8990271, 11352476,277242}, and public disclosure~\cite{liu2020vulnerabilitydistribution,bhandari2021cvefixes,pauley2023cve,wang2024reposvul}.
This design reduces dependence on any single reporting channel and 
provides a broader empirical basis for analyzing vulnerability structure, 
underlying failure mechanisms, and the coverage of existing discovery techniques.

\noindent
\textbf{(1) Linux Kernel Fixing Commits.}
Vulnerability-fixing commits from the Linux kernel mainline repository~\cite{linuxkernel} 
serve as the primary technical anchor of the dataset.
They provide direct evidence of how vulnerabilities are identified and resolved at the code level, 
including affected locations, triggering conditions, and remediation logic.
Because many kernel security issues are fixed without formal CVE assignment or public disclosure,
commit-level evidence is essential for capturing a broader set of real-world eBPF vulnerabilities 
beyond officially reported cases.
At the same time, such evidence is remediation-centered and may reflect maintainer practices,
patch granularity, and commit-message conventions.
These records therefore offer broad coverage and detailed technical context, 
but do not constitute a complete ground-truth view on their own.

\noindent
\textbf{(2) Syzbot Reports.}
syzbot~\cite{syzbot} is a continuous Linux kernel testing system built on syzkaller~\cite{syzkaller}.
Its reports provide discovery-side evidence, including crash types, triggering inputs, and fixing status~\cite{299527}.
Compared with fixing commits, which emphasize remediation,
syzbot captures how vulnerabilities are exposed in practice through automated testing.
This perspective is important because it provides concrete evidence about discovery conditions and 
fuzzing-relevant triggering behavior, supporting the later analysis of vulnerability discovery techniques.

\noindent
\textbf{(3) NVD / CVE Records.}
NVD~\cite{nvd} provides standardized public-disclosure records, including CVE identifiers,
vulnerability descriptions, and, in many cases, official CWE classifications.
Within our dataset, these records contribute a normalized and externally recognizable view of 
disclosed eBPF-related issues.
They complement commit- and report-level evidence by adding standardized identifiers,
disclosure metadata, and reference descriptions, thereby improving consistency for classification 
and cross-source validation.

Together, these sources combine remediation evidence, discovery-side evidence, 
and standardized disclosure records, providing the empirical foundation for the subsequent 
structural, mechanism-level, and technique-oriented analyses.

\subsection{Phase I: Data Collection and Integration} \label{sec:method:phase1}
Phase I constructs the empirical foundation of the study.
Its purpose is to collect candidate records related 
to real-world eBPF vulnerabilities from multiple sources, 
identify and validate the records to be retained through a unified evidence-guided procedure, 
and reconstruct them into a structurally consistent vulnerability dataset.

\subsubsection{Data Acquisition} \label{sec:method:phase1:collection}
This stage gathers candidate vulnerability records from the three data sources introduced in Section~\ref{sec:method:datasource}.
To ensure that the collected data reflect the mature and practically relevant evolution of the eBPF ecosystem, 
we use a unified time window spanning 2016 to 2025.
This period covers the stage in which eBPF entered the Linux mainline, matured with broader deployment, 
and overlapped with the active operational period of syzbot.
Using a common time window ensures that candidate records are drawn from a comparable and practically relevant period of eBPF development.

The acquisition process follows three design principles.
First, it adopts a \emph{coverage-first} strategy: candidate generation aims to capture potentially relevant instances broadly, 
rather than applying aggressive exclusion rules at the earliest stage.
Second, it follows a \emph{multi-signal consistency} principle: for each source, candidate records are identified by combining structured fields with textual or semantic cues, reducing dependence on any single heuristic.
Third, this stage is limited to candidate acquisition only.
It does not perform validity adjudication, cross-source deduplication, or final instance consolidation, thereby preserving an auditable raw candidate space for later integration.

For Linux kernel fixing commits~\cite{linuxkernel}, 
we treat vulnerability-fixing commits as candidate records and collect commit messages, 
modified file scopes, and compact code-diff information as locally persisted evidence.
To identify eBPF-related repair records, 
we apply a multi-stage filtering strategy combining subsystem path constraints with defect-repair semantic signals in commit messages.
We first restrict the search space to core eBPF-related directories to construct a subsystem-level candidate pool, 
and then filter out commits clearly unrelated to defect remediation based on message semantics.
This design preserves broad coverage while reducing irrelevant noise.
For NVD/CVE records~\cite{nvd}, 
we filter structured fields and textual descriptions to extract potentially eBPF-related entries.
For each retained record, we preserve vulnerability descriptions, CWE labels, and disclosure metadata.
For syzbot reports~\cite{syzbot}, 
we similarly filter structured fields and crash descriptions to extract eBPF-related fixed reports.
For each retained record, we preserve crash types, triggering inputs, and fixing status.

The result of this stage is a raw multi-source candidate set that serves as input to the subsequent cleaning, validation, and integration procedure.

\subsubsection{Data Cleaning}\label{sec:method:phase1:cleaning}

\begin{figure}[!ht]
\centering
\includegraphics[width=0.85\linewidth]{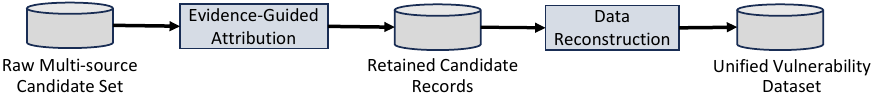}
\caption{Phase I workflow for transforming raw multi-source candidate records into the unified vulnerability dataset.}
\label{fig:core}
\end{figure}

After candidate acquisition, 
we apply an evidence-guided cleaning and integration procedure 
to determine which records should be retained in the study dataset and 
how they should be represented in a unified form.
As illustrated in Figure~\ref{fig:core}, 
this procedure first performs evidence-guided attribution over the raw multi-source candidate set, 
retains relevant records, and then reconstructs them 
into a unified vulnerability dataset with consistent structure and semantics.

\noindent
\textbf{(1) Evidence-Guided Attribution.}
This step determines which candidate records correspond to vulnerabilities rooted in the Linux kernel eBPF subsystem and should be retained for subsequent analysis.
To achieve this, 
we apply a unified evidence-guided attribution procedure consisting of four stages: 
automatic feature extraction, 
preliminary classification, 
manual adjudication, and retained record construction.
This design allows heterogeneous records from different sources to be evaluated under a common decision process while reducing reliance on any single heuristic signal.

\begin{table*}[!ht]
\caption{{\small Feature Categories and Evidence Signals for Core Attribution}}
\label{tab:attribution_features}
\centering
\small
\renewcommand{\arraystretch}{0.8}
\setlength{\tabcolsep}{4pt}
\begin{tabularx}{\linewidth}{lXX}
\toprule
\textbf{Category} & \textbf{Description} & \textbf{Role in Classification} \\
\midrule
Kernel Evidence
& Kernel crash/warning indicators (BUG, Oops, KASAN, etc.) and fix-related phrases (fixed in, regression, bisect)
& Determines whether a kernel-level fault is present; gates downstream rules \\
\midrule
BPF Core Code Signals
& Symbols and file paths belonging to the upstream eBPF subsystem (verifier, JIT, helpers, runtime)
& Identifies whether the fault location is within eBPF core code \\
\midrule
Call Trace Stack Analysis
& Positional analysis of top-$N$ and scan-$M$ frames from kernel call trace
& Attributes root cause by stack-top subsystem; distinguishes BPF-core vs. non-BPF faults \\
\midrule
Veto (Negative Evidence)
& Userspace lifecycle issues, tooling/framework bugs, allowed-semantics patterns
& Excludes non-kernel issues; priority $>$ positive evidence when no strong kernel signal \\
\midrule
Positive Upgrade
& Illegal execution context violations, persistent/irrecoverable failures
& Promotes entries to \texttt{core\_defect}/\texttt{core\_suspected} even without crash evidence \\
\midrule
Fix / Upstream Evidence
& Kernel commit links, upstream repo references, selftest mentions, regression hints
& Corroborates upstream kernel origin; upgrades confidence \\
\midrule
Semantic Value Signals
& Verifier soundness, JIT correctness, helper semantics, map concurrency, privilege/isolation
& Captures non-crash security-invariant violations in the eBPF subsystem \\
\midrule
Third-party Verifier
& Known non-upstream eBPF verifier/runtime names
& Strong negative: excludes non-upstream-kernel implementations \\
\bottomrule
\end{tabularx}
\end{table*}

\noindent
\textbf{\blackcircleone{A} Automatic Feature Extraction.}
For each candidate record, 
we derive a set of Boolean evidential signals from the available textual content.
For NVD entries, the input text includes vulnerability descriptions and reference-link pages.
For syzbot reports, it includes crash report titles, crash logs, and associated fixing information.
For kernel commits, it includes the commit subject and body, the list of modified files, and a compact diff summary.
All features are extracted using deterministic rules to ensure reproducibility.
Table~\ref{tab:attribution_features} summarizes the feature categories and their roles in the attribution process.
By transforming heterogeneous textual evidence into a structured representation, this step provides a consistent basis for downstream screening and adjudication.

\noindent
\textbf{\blackcircleone{B} Preliminary Classification.}
Using the extracted evidential signals, we apply automated scripts to determine each record’s relevance to the Linux kernel eBPF subsystem.
At this stage, records are assigned two primary labels, \texttt{core\_defect} and \texttt{non\_core}, together with an auxiliary label, \texttt{core\_suspected}, for cases whose evidence is incomplete but potentially relevant to the core subsystem.
This stage provides scalable filtering while preserving uncertain cases for further inspection rather than excluding them prematurely.

\noindent
\textbf{\blackcircleone{C} Manual Adjudication.}
All records labeled \texttt{core\_suspected} are manually reviewed based on the full available evidence.
A record is retained as \texttt{core\_defect} if it involves both kernel-side fix behavior and eBPF-related code paths; otherwise it is labeled \texttt{non\_core}.
To improve reliability, 
all records requiring manual labeling are independently annotated by two researchers, 
and disagreements are resolved through discussion until consensus is reached.
This step improves the robustness of the attribution process for ambiguous cases that cannot be resolved confidently through automated rules alone.

\noindent
\textbf{\blackcircleone{D} Retained Record Construction.}
After attribution is completed, we retain two categories of records:
(i) all \texttt{core\_defect} records, and
(ii) \texttt{non\_core} records corresponding to officially maintained user-space tools such as \texttt{libbpf} and \texttt{bpftool}.
The latter are retained because they are part of the officially maintained eBPF ecosystem and remain closely tied to the kernel subsystem in both functionality and security evolution.
This retained set forms the input to the subsequent reconstruction stage.

To assess the quality of this procedure, 
we perform three validation steps using random sampling over the corresponding decision populations~\cite{lohr2021sampling}.
First, we randomly sample 600 records from the Stage~B automated decisions for manual verification.
The sample size is determined using standard proportion-estimation with a 95\% confidence level, 
a 4\% margin of error, and a conservative proportion setting of $p=0.5$.
The resulting verification accuracy is 94.3\%.
Second, we randomly sample 600 records from the final retained dataset for independent verification, 
yielding an accuracy of 96.7\%.
Third, we randomly sample 200 kernel commits to examine potential duplication at the commit level, 
with an observed duplication rate of 5\%.
Together, 
these validation steps assess the accuracy of automated attribution, 
the reliability of the final dataset, and the degree of residual duplication in the commit-based representation.
The results indicate that the attribution procedure provides a reliable basis 
for dataset construction and subsequent analyses, 
while acknowledging residual fragmentation across related records.

\vspace{3pt}
\noindent
\textbf{(2) Data Reconstruction.}\label{sec:method:phase1:cleaning:reconstruction}
After the retained records are finalized, 
we reconstruct them into a unified vulnerability dataset for subsequent analysis.
We first organize each retained record 
using the unified issue representation shown in Table~\ref{tab:issue-structure}.
We then perform cross-source linking by extracting commit hashes from CVE and syzbot records and 
associating them with the corresponding kernel commits~\cite{akhoundali2024morefixes}.
In subsequent analyses, 
multiple records linked to the same commit are counted as a single instance, 
while records that cannot be linked are treated as independent instances.
This rule reduces straightforward cross-source duplication, 
but cannot fully eliminate residual fragmentation when one underlying issue spans multiple related commits 
or when source records lack reliable linking metadata.
This reconstruction step integrates heterogeneous evidence 
from vulnerability discovery, disclosure, and remediation into a unified dataset 
while preserving the multi-dimensional information associated with each case.
The resulting unified vulnerability dataset serves as the foundation 
for the classification, root-cause analysis, and technique assessment stages that follow.

\begin{table}[htp]
\caption{Structure of the Unified Issue Representation.}
\label{tab:issue-structure}
\centering
\small
\renewcommand{\arraystretch}{1.0}
\setlength{\tabcolsep}{4pt}
\begin{tabularx}{0.6\linewidth}{c l X l}
\toprule
\textbf{No.} & \textbf{Field} & \textbf{Sub-field} & \textbf{Description} \\
\midrule
1 & \texttt{uid} & --- & Unique issue identifier \\
\cline{1-4}
2 & \multirow{3}{*}{\texttt{issue}} & \texttt{BPF-Component} & \textit{Verifier / JIT / Runtime / Userspace} \\
3 &  & \texttt{Root Cause} & Provide a concise description of each case \\
4 &  & \texttt{Essence} & Abstract of the fundamental flaw \\
\bottomrule
\end{tabularx}
\end{table}

\subsection{Phase II: CWE-Based Classification}\label{sec:method:phase2}

Phase II organizes the unified vulnerability dataset 
into a consistent hierarchical classification framework for subsequent analysis.
To achieve this, 
we adopt the MITRE-provided \emph{CWE-1000: Research Concepts View}~\cite{cwe1000} and 
perform CWE annotation and hierarchical classification following MITRE mapping guidance~\cite{cwe_mapping_guidance}.
Compared with other CWE views, 
CWE-1000 is better suited to this study because it is designed for research-oriented analysis, 
provides a hierarchical organization of weakness concepts, 
and supports consistent comparison across multiple abstraction levels.
The result of this phase is a categorized dataset that 
serves as the basis for the structural and mechanism-level analyses.

\subsubsection{CWE Annotation}\label{sec:method:phase2:annotation}
This step assigns a standardized CWE label to each vulnerability record in the unified dataset.
Its purpose is to establish a consistent weakness-level representation 
that supports hierarchical classification and cross-record comparison.

\vspace{3pt}
\noindent
\textbf{(1) Annotation Scope and Method.}
The CWE-1000 view contains four hierarchical levels of entries: Variant, Base, Class, and Pillar.
According to MITRE guidance, Base- or Variant-level mappings should be preferred whenever possible, 
while higher-level categories should be used only when lower-level assignments are not appropriate.
In practice, however, Variant-level entries are often overly fine-grained for population-level analysis, 
whereas Base-level entries typically provide sufficient semantic precision 
while maintaining stronger statistical stability.
Accordingly, we prioritize Base-level assignments whenever possible and fall back 
to higher levels (Class or Pillar) only when no suitable Base category exists.
For records without an official CWE identifier, 
we apply this annotation strategy directly.
For records with existing CWE assignments, 
we normalize non-Base labels along the CWE-1000 hierarchy.
Specifically, 
if the official label is at the Variant level, 
we map it to a semantically appropriate Base-level parent; 
if it is at the Class or Pillar level, 
we map it to a clearly appropriate Base-level child.
Such reassignment is performed only when the hierarchy provides a clear and semantically consistent correspondence.
This strategy improves comparability across records 
while preserving the underlying weakness semantics.

To ensure annotation quality, 
we employ \emph{Dual Independent Annotation with Consensus Resolution}~\cite{cohen1960kappa,fleiss1971kappa} 
together with \emph{blind assessment}~\cite{neuendorf2002content}.
For all records requiring manual annotation, 
two researchers independently assign CWE labels based on the full record evidence, 
and disagreements are resolved through discussion until consensus is reached.
This procedure reduces annotator bias and improves assignment reliability.
In addition, for CVE records with official CWE identifiers, 
the original labels are temporarily hidden during validation, 
and the same annotation procedure is applied without access to them.
The resulting manual assignments are then compared against the hidden official labels.
This blind assessment provides an external consistency check 
and evaluates alignment with standardized CWE mappings.

\vspace{3pt}
\noindent
\textbf{(2) Quality Validation.}
The blind assessment yields an agreement rate of 81\%.
Further inspection shows that most disagreements occur between adjacent or 
closely related CWE categories in the hierarchy,
indicating differences in granularity rather than fundamentally inconsistent interpretations.
Accordingly, the assigned CWE labels should be interpreted 
as a standardized analytical representation for cross-record comparison 
rather than exact ground-truth semantics for individual cases.
In particular, some borderline instances may reasonably admit adjacent mappings 
within the CWE hierarchy without materially affecting the overall structural conclusions.

\subsubsection{Hierarchical Classification}\label{sec:method:phase2:cwe_classification}

After completing CWE annotation, 
we organize the dataset according to each assigned CWE identifier’s position in the CWE-1000 hierarchy.
This step transforms record-level assignments into a structured hierarchical representation 
that supports comparison across different abstraction levels.
Specifically, 
we construct three complementary views:
(i) a Base-level distribution,
(ii) a Class--Base two-level distribution, and
(iii) a Pillar--Class--Base three-level distribution.
These views preserve both fine-grained weakness types and their aggregation relationships within the CWE taxonomy.

To support subsequent component-oriented analysis, 
we further partition the dataset by eBPF component and 
apply the same hierarchical classification to each subset.
This yields both a global hierarchical view and component-specific views under a common framework.
As a result, 
the dataset can be examined not only in terms of overall weakness distribution, 
but also in terms of how different vulnerability classes are distributed across major eBPF components.
Together, 
these classification results establish a common analytical framework for later phases.
They enable structural concentration analysis across abstraction levels, 
support identification of dominant weakness patterns, and 
provide a consistent basis for cross-component comparison.
\subsection{Phase III: Empirical Analysis}

This phase conducts the main empirical analyses of the study 
using the categorized vulnerability dataset produced in Phase II.
Its purpose is to characterize structural concentration in the real-world eBPF vulnerability landscape, 
identify dominant underlying failure mechanisms, 
analyze how these mechanisms are distributed across major eBPF components and execution stages, 
and assess how well existing representative techniques cover the resulting vulnerability patterns.

\subsubsection{Statistical Analysis}

This step analyzes the full categorized dataset 
to characterize the structural concentration of real-world eBPF vulnerabilities 
across multiple abstraction levels.
It serves as the starting point of Phase III by identifying dominant weakness patterns 
and establishing the empirical basis for subsequent mechanism-level analysis.
To do so, 
we examine the dataset through three complementary hierarchical views: 
the Base-level distribution, the Class--Base two-level distribution, and the Pillar--Class--Base three-level distribution.
The Base-level view captures common fine-grained weakness types, 
while higher-level views reveal how these weaknesses aggregate into broader structural categories.
Together, 
these views enable analysis at different abstraction levels 
while preserving hierarchical relationships among weakness types.
Our analysis focuses on three aspects:
\begin{itemize}
\item the frequency distribution and concentration of CWE categories across abstraction levels;
\item the cumulative coverage of high-frequency categories, 
to assess whether the vulnerability landscape exhibits a head-heavy or long-tail structure; and
\item differences in vulnerability-type distributions across major eBPF components, 
to support subsequent mechanism-level and architectural analyses.
\end{itemize}

\subsubsection{Mechanism-Level Analysis}\label{sec:sampling}

This step identifies the dominant failure mechanisms underlying the real-world eBPF vulnerability landscape.
Because mechanism-level root-cause analysis requires manual inspection of traceable evidence 
and cannot be reliably performed at scale over the full dataset, 
we construct a reproducible analytical subset using stratified sampling 
and conduct in-depth analysis on that subset.
This design allows the study to move beyond category-level distributions 
while preserving structural representativeness of the sampled data~\cite{Cochran1977,lohr2021sampling}.

\vspace{3pt}
\noindent
\textbf{(1) Stratification Design and Sample Size Determination.}
We adopt the Class level of CWE-1000 as the stratification dimension.
This level provides a balance between semantic discriminability and statistical stability: 
it is more specific than the Pillar level, which aggregates conceptually distinct weaknesses into broad groups, 
while avoiding the fragmentation that would arise at the Base or Variant levels, 
where many categories are too fine-grained to support stable sampling and comparison.
Using the Class level therefore preserves meaningful structural differences 
without introducing excessive sparsity.
For the small number of Base CWEs without an intermediate Class ancestor, 
we aggregate them under their associated Pillar to ensure all strata remain mutually exclusive and collectively exhaustive.

To reduce instability caused by extremely small strata, 
we include only strata with at least 100 instances in the root-cause sampling.
This threshold avoids drawing conclusions from sparsely populated strata,
where random variation would dominate.
The included strata account for 84.7\% of the total population, 
indicating that the sampled analysis covers the dominant structural portion of the dataset while limiting noise from very-low-frequency categories.
Excluded strata remain in the full dataset for distributional analysis; 
their omission affects only the mechanism-level analysis.
Accordingly, 
the mechanism-level findings explain the dominant portion of the vulnerability landscape rather than 
exhaustively characterize all low-frequency categories.

Following classical sample-size determination for proportion estimation~\cite{Cochran1977}, 
we compute a target sample size of approximately 600 under a 95\% confidence level with a 4\% margin of error.
This provides a statistically grounded size for constructing a manually analyzable subset while maintaining coverage of dominant strata.
For all included strata, we apply proportional allocation~\cite{Cochran1977,lohr2021sampling} with a per-stratum minimum sample floor:
\begin{equation}
  n_{\min} = 50
\end{equation}
to prevent within-stratum samples from becoming too small for reliable mechanism identification.
Without this floor, smaller but important strata could contribute too few samples for stable qualitative analysis.
The allocation formula is:
\begin{equation}
  n_i = \max\!\left(\mathrm{round}\!\left(\frac{N_i}{N} \cdot n\right),\; n_{\min}\right)
\end{equation}
where $N_i$ is the size of the $i$-th stratum, $N = 2{,}342$ is the total size of included strata, 
and $n = 600$ is the target sample size.
Due to rounding and the minimum floor, the final sample size is 642.

\vspace{3pt}
\noindent
\textbf{(2) Sampling Execution and Root-Cause Analysis Procedure.}
\label{sec:sampling-execution}
Within each selected stratum, 
we perform simple random sampling without replacement~\cite{Cochran1977} 
using a fixed random seed (\texttt{seed\,=\,42}) to ensure reproducibility.
For each sampled instance, 
we assign a mechanism label together with its associated eBPF component and execution-stage location 
based on the structured representation and traceable evidence constructed in Phase I.
To assess labeling consistency, 
a subset of sampled instances is independently labeled by two authors, 
with disagreements resolved through discussion.
We then aggregate these labels and interpret the resulting mechanism-level patterns 
alongside the full dataset’s hierarchical CWE and component-level distributions.
The sampled subset provides mechanism-level detail, while the full dataset preserves population-level context 
for architectural analysis (RQ3) and technique assessment.

\subsubsection{Technique Assessment}

This step evaluates how well existing techniques represent the dominant real-world eBPF vulnerability patterns 
identified in the preceding analyses.
Although userspace issues are retained in the unified dataset for landscape analysis, 
the technique assessment is scoped to kernel-side discovery capability.

\vspace{3pt}
\noindent
\textbf{(1) Evaluated Techniques and Selection Rationale.}
Real-world eBPF vulnerabilities often involve runtime states, 
cross-component interactions, and semantic constraints that are difficult to exercise systematically~\cite{281444,291291,hao2022dependencychallenge}.
Among existing approaches, 
fuzzing-based dynamic techniques are particularly relevant 
because they explore concrete execution paths under realistic system states 
and are widely used for vulnerability discovery in the Linux kernel and eBPF ecosystem.
Since many dominant vulnerability patterns depend on 
passing verification, constructing valid execution contexts, 
and triggering deep runtime behavior, 
dynamic fuzzing-based techniques provide an appropriate basis for assessing practical discovery coverage.
Accordingly, this study focuses on representative fuzzing-based techniques.
Specifically, 
we analyze \textit{Syzkaller}~\cite{syzkaller}, \textit{Buzzer}~\cite{google_buzzer}, and \textit{BRF}~\cite{hung2024brf}.
\textit{Syzkaller} is a general-purpose coverage-guided kernel fuzzer and 
serves as the core infrastructure of \textit{Syzbot}~\cite{shi2019enterprisefuzzing}.
\textit{Buzzer} focuses on verifier-oriented exploration, 
whereas \textit{BRF} aims to improve verifier pass rates 
and increase reachability into the Runtime stage.
We characterize their capability boundaries along three dimensions: 
input generation, state construction, and semantic modeling~\cite{280702}.

\vspace{3pt}
\noindent
\textbf{(2) Empirical Evaluation Procedure.}
We first conduct a design-level analysis of the selected techniques 
based on their publications, documentation, and architectural design.
This analysis focuses on input-generation strategies~\cite{217573,hao2025syzspec}, state-construction capabilities, 
path-exploration mechanisms, and modeling of key eBPF semantics.
Using this analysis, 
we assess design-level coverage potential across mechanism categories identified earlier, 
along two dimensions: construction capability and detection capability.
Building on this analysis, 
we select \textit{Syzkaller} as the platform for empirical evaluation.
This choice is motivated by three considerations:
its broad coverage across stages from program loading to runtime interaction;
its mature and reproducible execution environment; and
its suitability for evaluating general-purpose dynamic techniques rather than specialized tools.

For empirical evaluation, 
we select Linux v5.10 LTS as the target kernel version 
because it contains the largest concentration of relevant vulnerability instances in our dataset 
and provides a practical common basis for replay and comparison.
This choice supports a focused empirical analysis 
but does not imply that v5.10 is uniquely representative of all kernel versions.
Accordingly, 
the evaluation is intended to illustrate the observed fuzzing gap rather than to estimate a version-invariant population parameter.
We deploy 8 parallel QEMU virtual machines, each configured with 2 CPU cores and 3\,GB memory, 
running 8 fuzzer processes per VM for a continuous 72-hour campaign.
During the experiment, 
we collect coverage curves, per-file coverage breakdowns, and crash/sanitizer reports.
We further perform systematic replay analysis of all corpus seeds generated by \textit{Syzkaller} 
to decompose aggregate coverage into fine-grained signals:
\begin{itemize}
\item \textit{Seed structure parsing.} Extract program type, map types, helper calls, and attach mechanisms.
\item \textit{Coverage replay.} Re-execute each seed to collect KCOV coverage and verifier outcomes.
\item \textit{Coverage attribution.} Attribute coverage to Pass / Reject / No-Load categories.
\item \textit{Semantic diversity analysis.} Measure how many distinct \texttt{prog\_type}s reach each covered PC.
\item \textit{Crash--seed matching.} Link crash reports to triggering seeds.
\end{itemize}

Together, the cross-tool analysis and the \textit{Syzkaller}-based case study 
identify coverage and observability gaps between current discovery techniques 
and dominant real-world vulnerability patterns.

\section{Empirical Results} \label{sec:empirical}

In this section, 
we present the empirical findings of our study following the staged methodology described in Section~4. 
We begin with an overview of the constructed dataset, 
and then present results corresponding to each research question (RQ1–RQ4), 
moving from structural characterization to mechanism-level analysis, 
architectural interpretation, and technique-level assessment.

\subsection{Data Overview}
\label{sec:results:overview}

We initially collected 11,388 raw records from three data sources and 
processed them through the cleaning, attribution, cross-source linking, 
and classification procedures described in Section~\ref{sec:method}. 
Table~\ref{tab:dataset-composition} summarizes the final curated dataset, which contains 2,766 vulnerability instances.

\begin{table}[!ht]
{
\centering
\small
\renewcommand{\arraystretch}{0.8}
\setlength{\tabcolsep}{5pt}
\begin{minipage}[t]{0.49\linewidth}
\centering
\captionof{table}{{\small Source composition of the final dataset. 
Counts and percentages are reported over the 2{,}766 vulnerability instances.}}
\label{tab:dataset-composition}
\begin{tabular}{lrr}
\toprule
\textbf{Source} & \textbf{Count} & \textbf{Percentage} \\
\midrule
Kernel Commits  & 2{,}439 & 88.2\% \\
NVD (CVE)       &   197   &  7.1\% \\
Syzbot          &   130   &  4.7\% \\
\midrule
\textbf{Total}  & \textbf{2{,}766} & \textbf{100\%} \\
\bottomrule
\end{tabular}
\end{minipage}
\hfill
\begin{minipage}[t]{0.49\linewidth}
\centering
\captionof{table}{{\small CWE distribution across abstraction levels,
\#CWEs denotes categories, and \#Entries denotes vulnerability instances.}}
\label{tab:cwe-level-dist}
\begin{tabular}{lrr}
\toprule
\textbf{Abstraction Level} & \textbf{\#CWEs} & \textbf{\#Entries} \\
\midrule
Base    & 25 & 1{,}840 \\
Class   & 10 &   744 \\
Pillar  &  3 &   182 \\
\midrule
\textbf{Total} & \textbf{38} & \textbf{2{,}766} \\
\bottomrule
\end{tabular}
\end{minipage}
}
\end{table}

\begin{table}[!ht]
{
\caption{{\small Global Pillar$\rightarrow$Class$\rightarrow$Base hierarchy of the dataset. \#Base denotes Base-level CWE categories aggregated under each branch, and \#Entries denotes vulnerability instances.}}
\label{tab:cwe_hierarchy}
\centering
\small                
\renewcommand{\arraystretch}{0.8} 
\setlength{\tabcolsep}{4pt}  
\begin{tabularx}{0.9\linewidth}{XXlll}
\toprule
\textbf{Pillar} & \textbf{Class} & \textbf{\#Base} & \textbf{\#Entries} & \textbf{\%} \\
\midrule
\multirow{9}{*}{\makecell[l]{CWE-664\textit{ (Resource Lifetime)}}}
  & CWE-119 (Memory Bounds)        & 6 & 735 & 26.6 \\
  & CWE-362 (Race Condition)       & 3 & 331 & 12.0 \\
  & CWE-404 (Resource Release)     & 1 & 198 &  7.2 \\
  & CWE-665 (Initialization)       & 2 & 166 &  6.0 \\
  & CWE-667 (Locking)              & 2 & 156 &  5.6 \\
  & CWE-704 (Type Conversion)      & 3 & 137 &  5.0 \\
  & CWE-400 (Resource Consumption) & 2 &  91 &  3.3 \\
  & CWE-672 (Use After Release)    & 1 &  68 &  2.5 \\
  & CWE-200 (Info Exposure)        & 2 &  48 &  1.7 \\
\midrule
\makecell[l]{CWE-710\textit{ (Coding Standards)}}
  & ---                              & 1 & 256 & 9.3 \\
\makecell[l]{CWE-682\textit{ (Incorrect Calc.)}}
  & ---                              & 4 & 202 & 7.3 \\
\makecell[l]{CWE-707\textit{ (Neutralization)}}
  & CWE-20 (Input Validation)        & 4 & 161 & 5.8 \\
\makecell[l]{CWE-703\textit{ (Exception Handling)}}
  & CWE-754 (Unusual Conditions)     & 3 &  77 & 2.8 \\
\multirow{2}{*}{\makecell[l]{CWE-691\textit{ (Control Flow)}}}
  & CWE-834 (Excessive Iteration)    & 1 &  55 & 2.0 \\
  & CWE-674 (Uncontrolled Recursion) & 1 &  12 & 0.4 \\
\makecell[l]{CWE-284\textit{ (Access Control)}}
  & ---                              & 1 &  45 & 1.6 \\
\midrule
\multicolumn{2}{l}{\textbf{Total (7 Pillars, 13 Classes, 38 CWEs)}}
  & \textbf{---} & \textbf{2{,}766} & \textbf{100} \\
\bottomrule
\end{tabularx}
}
\end{table}

\begin{table}[!ht]
\caption{{\small Component-wise distribution of the dataset across CWE abstraction levels.
For each component, \#CWE denotes the number of distinct categories, 
and \#Entry denotes the number of vulnerability instances assigned at that abstraction level.
The Base/Class/Pillar columns therefore report level-specific counts rather than ancestor-expanded repeats.}}
\label{tab:cwe-level-dist-component}
\centering
\small                
\renewcommand{\arraystretch}{0.8} 
\setlength{\tabcolsep}{4pt}  
\begin{tabularx}{0.9\linewidth}{l XX XX XX XX}
\toprule
 & \multicolumn{2}{c}{\textbf{Base}} & \multicolumn{2}{c}{\textbf{Class}} & \multicolumn{2}{c}{\textbf{Pillar}} & \multicolumn{2}{c}{\textbf{Total}} \\
\cmidrule(lr){2-3} \cmidrule(lr){4-5} \cmidrule(lr){6-7} \cmidrule(lr){8-9}
\textbf{Component} & \#CWE & \#Entry & \#CWE & \#Entry & \#CWE & \#Entry & \#CWE & \#Entry \\
\midrule
Runtime     & 25 & 1{,}209 & 10 & 577 & 3 &  65 & 38 & 1{,}851 \\
Verifier    & 25 &   335   & 10 & 118 & 3 &  67 & 38 &   520 \\
JIT         & 22 &   145   &  8 &  27 & 2 &  50 & 32 &   222 \\
Userspace   & 17 &   150   &  9 &  22 & 0 &   0 & 26 &   172 \\
\midrule
\textbf{Overall} & \textbf{25} & \textbf{1{,}840} & \textbf{10} & \textbf{744} & \textbf{3} & \textbf{182} & \textbf{38} & \textbf{2{,}766} \\
\bottomrule
\end{tabularx}
\end{table}

As shown in Table~\ref{tab:dataset-composition}, 
the final curated dataset contains 2,766 vulnerability instances, 
of which kernel fixing commits contribute 2,439 (88.2\%), 
while NVD/CVE records and syzbot reports contribute 197 (7.1\%) and 130 (4.7\%), respectively. 
This source composition indicates that public disclosure and automated reporting alone cover only a limited portion of the eBPF vulnerability landscape, 
which motivates the multi-source dataset construction adopted in this study.
Because most retained instances come from kernel fixing commits, 
the curated dataset should also be understood as a remediation-centered empirical view, 
which may inherit recording biases from commit practices.

To support the subsequent analyses, 
we further organize the dataset under the CWE-1000 hierarchy at three abstraction levels. 
Table~\ref{tab:cwe-level-dist} summarizes the overall distribution across these levels. 
In total, the dataset spans 38 CWE categories, including 25 Base-level categories, 10 Class-level categories, and 3 Pillar-level categories. 
Among the 2,766 instances, 1,840 are represented at the Base level, 744 at the Class level, and 182 at the Pillar level.
Table~\ref{tab:cwe_hierarchy} further summarizes how the curated dataset is organized under the global Pillar$\rightarrow$Class$\rightarrow$Base hierarchy. 
Each row corresponds to one hierarchy branch and reports both the number of represented Base-level categories and the number of vulnerability instances mapped to that branch. 
For example, under the Pillar CWE-664 (Resource Lifetime), the branch CWE-119 (Memory Bounds) contains 6 Base-level categories and 735 instances, while CWE-362 (Race Condition) contains 3 Base-level categories and 331 instances. 
This table provides a global view of how the dataset is distributed across major hierarchy branches.

Beyond the global hierarchy, 
we also organize the dataset by major eBPF components: Runtime, Verifier, JIT, and Userspace. 
Table~\ref{tab:cwe-level-dist-component} reports, for each component, 
the numbers of vulnerability instances assigned at the Base, Class, and Pillar levels, respectively. 
These counts are grouped by the final abstraction level of each instance’s CWE assignment, 
rather than by propagating each instance upward to all ancestor levels in the hierarchy. 
Thus, 
the Base/Class/Pillar columns should be interpreted as a partition of the dataset 
by assigned abstraction level within each component. 
For example, 
Runtime contains 1,209 instances assigned at the Base level, 
577 assigned at the Class level, and 65 assigned at the Pillar level, 
for a total of 1,851 Runtime instances. 
This component-wise view complements the global hierarchy by showing 
how the curated dataset is distributed across the major parts of the eBPF ecosystem. 
While the overall dataset includes both kernel-core and closely related userspace-ecosystem issues, 
the later architectural interpretation emphasizes kernel-side risk-bearing roles.

\begin{tcolorbox}[colback=gray!15,colframe=gray!60,boxrule=0.8pt,arc=4pt]
\textbf{TA 1.} As shown in Table~\ref{tab:dataset-composition}, 
88.2\% of the curated instances come from kernel fixing commits, 
whereas publicly disclosed CVE entries and syzbot reports together account for only about 12\%. 
This suggests that public disclosure and automated reporting alone 
do not provide a sufficiently complete view of the eBPF vulnerability landscape, 
motivating the multi-source dataset construction adopted in this study.
\end{tcolorbox}

\subsection{RQ1: What structural concentration patterns characterize real-world eBPF vulnerabilities?}

\subsubsection{Base-level distribution.}

\begin{figure}[!ht]
\centering
\includegraphics[width=\linewidth]{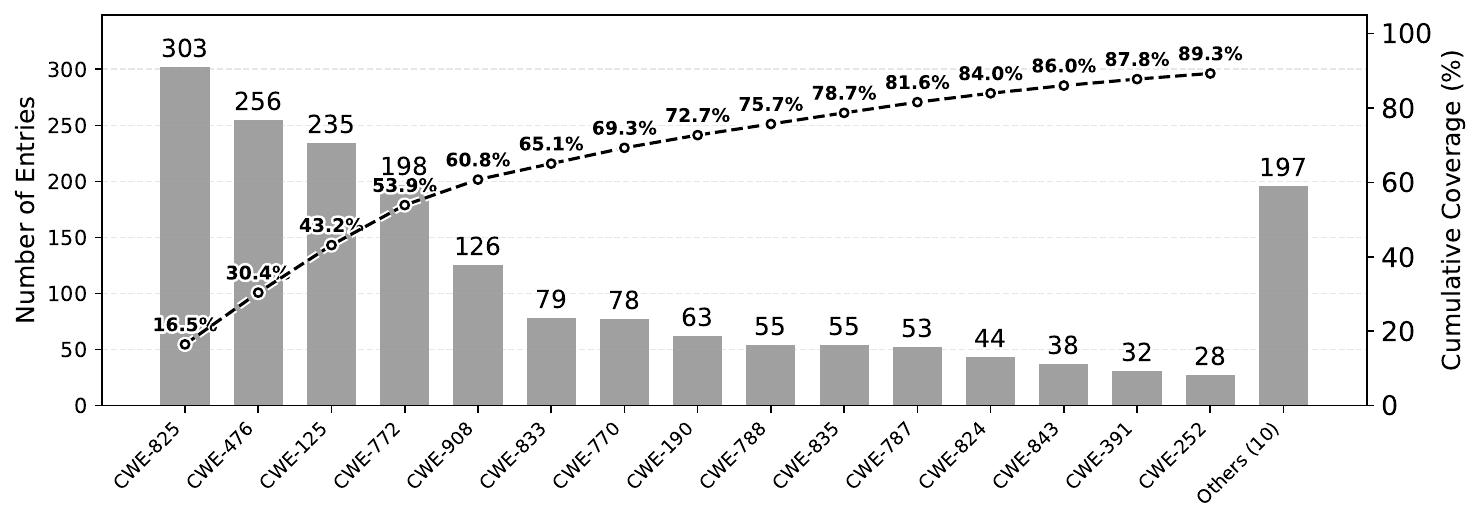}
\caption{Top-15 Base-level CWE categories ranked by frequency, with cumulative coverage shown by the line. 
The figure illustrates a clear heavy-tail pattern, in which a small number of categories account for most Base-level instances.}
\label{fig:base_topk}
\end{figure}

At the finest level of abstraction, 
Figure~\ref{fig:base_topk} presents the most frequent Base-level CWEs together with their cumulative coverage. 
The results show a clear \textit{heavy-tail distribution}: 
a small number of weakness types account for most observed vulnerability instances, 
while the remaining categories each contribute relatively small counts. 
In particular, 
the most frequent Base-level categories include 
\textit{CWE-825} (303 instances), \textit{CWE-476} (256), \textit{CWE-125} (235), and \textit{CWE-772} (198). 
Overall, 
the top five Base CWEs cover \textbf{60.8\%} of all Base-level instances, 
and the top fifteen cumulatively account for 89.3\%, 
leaving only 10.7\% distributed across the remaining categories.

Several patterns emerge from this distribution. 
First, 
the concentration is already pronounced at the most fine-grained level, 
indicating that the observed landscape is not broadly dispersed across many unrelated Base-level weakness types. 
Second, 
the head of the distribution is largely composed of categories related to \textit{memory safety}, 
\textit{resource lifetime management}, and \textit{concurrency-related defects}, 
including out-of-bounds accesses (\textit{CWE-125}), null-pointer dereferences (\textit{CWE-476}), 
improper resource release (\textit{CWE-825}, \textit{CWE-772}), 
and lock- or blocking-related issues such as \textit{CWE-833}. 
Third, 
the cumulative curve rises rapidly over the first few categories and then flattens substantially, 
showing that most of the structural mass of the Base-level landscape 
is concentrated in a relatively small prefix of categories.

Because some instances are annotated only at the Class or Pillar level, 
the Base-level view primarily characterizes dominant fine-grained defect patterns rather than the full population. 
Even so, 
it provides the most detailed view of the recurring vulnerability categories that dominate the observed landscape.

\subsubsection{Class-level distribution.}

\begin{figure}[!ht]
\centering
\includegraphics[width=\linewidth]{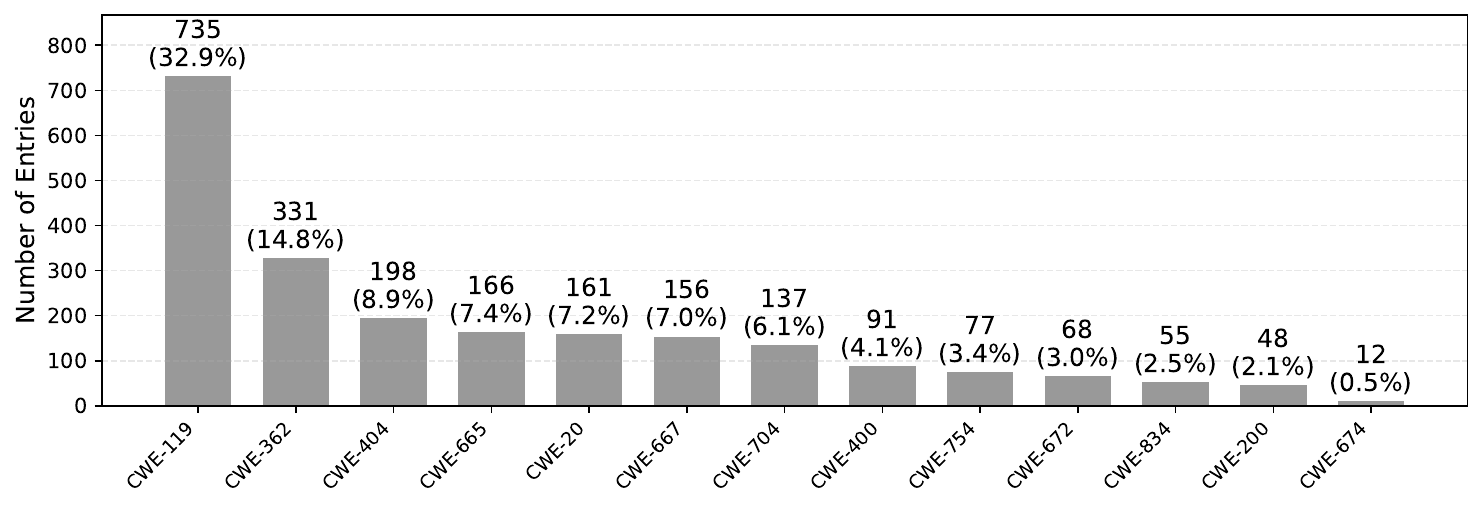}
\caption{Class-level CWE distribution of the dataset. 
The figure shows that the concentration observed at the Base level remains strong after aggregation into broader weakness groups.}
\label{fig:class_dis}
\end{figure}

Building on the Base-level results, 
Figure~\ref{fig:class_dis} shows the distribution of vulnerabilities at the Class level. 
Because this level includes both instances directly annotated at the Class level and 
those aggregated upward from Base nodes, 
it provides a clearer view of whether the concentration observed above persists 
after related fine-grained weakness types are merged into broader groups. 
The results show that the distribution remains strongly imbalanced. 
In particular, 
\textit{CWE-119 (Memory Bounds)} alone accounts for \textbf{32.9\%} of all vulnerability instances, 
while \textit{CWE-362 (Race Condition)} contributes an additional 14.8\%. 
Together, 
these two categories account for 47.7\% of the full dataset. 
In addition, 
\textit{CWE-404 (Resource Release Errors)}, \textit{CWE-665 (Initialization Errors)}, and \textit{CWE-667 (Locking-Related Issues)} also appear with relatively high frequencies.

Compared with the Base-level view, 
the Class-level aggregation makes the dominant structural themes clearer. 
The strong skew remains after related weakness types are merged into broader categories, 
indicating that the observed concentration is not simply an artifact of fine-grained labeling. 
Instead, 
the landscape remains organized around a limited number of broader weakness groups, 
particularly memory bounds, concurrency, resource release, initialization, and locking-related issues.

\subsubsection{Pillar-level distribution.}

\begin{figure}[!ht]
\centering
\includegraphics[width=\linewidth]{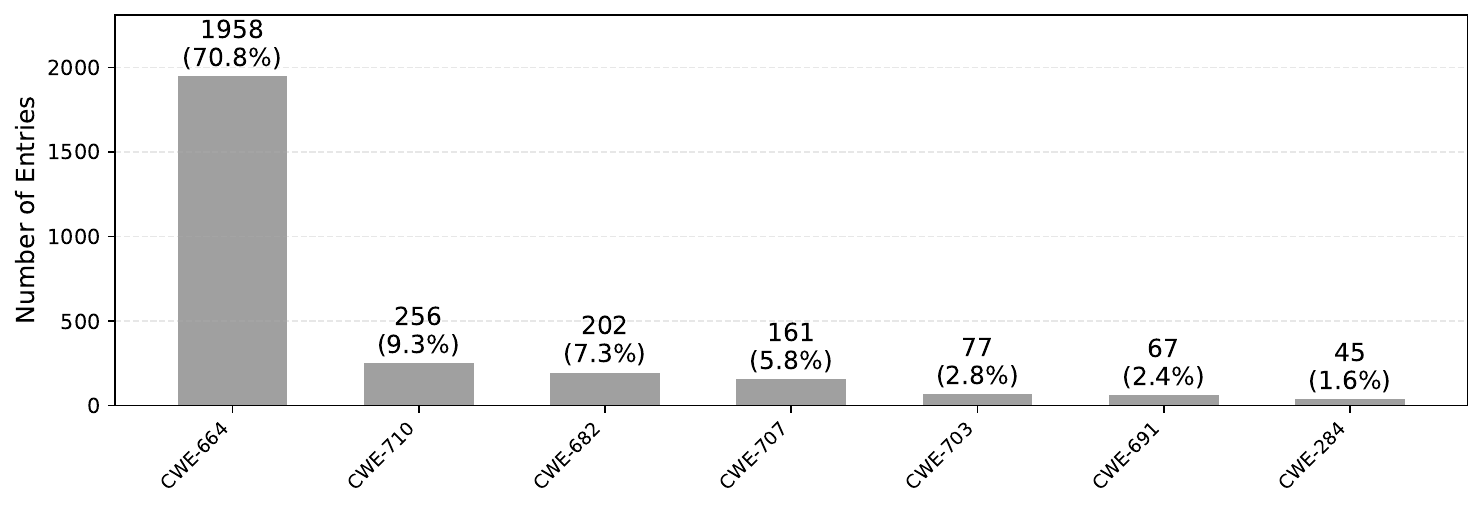}
\caption{Pillar-level CWE distribution of the dataset. 
The figure shows that the structural concentration observed at the Base and Class levels becomes even stronger at the highest level of aggregation.}
\label{fig:pillar_dis}
\end{figure}

At the highest level of abstraction, 
Figure~\ref{fig:pillar_dis} reveals an even stronger concentration pattern. 
\textit{CWE-664 (Resource Lifetime)} alone accounts for \textbf{70.8\%} of all vulnerability instances, 
while the remaining Pillar categories each contribute comparatively small proportions. 
This shows that the concentration observed at the Base and Class levels persists at the broadest level and becomes even more pronounced after aggregation. 
At this level, 
the dominant portion of the real-world eBPF vulnerability landscape is concentrated in a very small set of broad weakness families, 
with resource lifetime and system-state consistency issues occupying the central position.

\begin{tcolorbox}[colback=gray!15,colframe=gray!60,boxrule=0.8pt,arc=4pt]
\textbf{TA 2. Across the observed dataset, 
eBPF vulnerabilities exhibit a clear and persistent structural concentration pattern. 
Rather than being broadly dispersed, the observed landscape is dominated by a relatively small number of high-frequency weakness categories and their aggregated groups. 
This answers RQ1 and motivates the deeper mechanism-level analysis in RQ2.}
\end{tcolorbox}

\subsection{RQ2: What mechanism-level failure modes underlie the dominant real-world eBPF vulnerabilities?}
\label{sec:rq2}

To answer RQ2, 
we move beyond the structural concentration patterns identified in RQ1 
and examine dominant vulnerabilities at the mechanism level 
using a stratified sampled subset that preserves the major structural strata of the dataset. 
Table~\ref{tab:sample-hierarchy} summarizes the sampled subset used for this analysis. 
Based on this subset, 
we perform detailed root-cause analysis to identify the system-level failure modes 
that underlie the dominant portion of the real-world eBPF vulnerability landscape.

\begin{table}[!ht]
\caption{{\small CWE hierarchy of the stratified root-cause sample. 
\#Base denotes represented Base-level CWE categories, 
and \#Entries denotes sampled vulnerability instances.}}
\label{tab:sample-hierarchy}
\centering
\small                
\renewcommand{\arraystretch}{0.8} 
\setlength{\tabcolsep}{4pt}  
\begin{tabularx}{0.9\linewidth}{XXrrr}
\toprule
\textbf{Pillar} & \textbf{Class} & \textbf{\#Base} & \textbf{\#Entries} & \textbf{\%} \\
\midrule
\multirow{6}{*}{\makecell[l]{CWE-664\textit{ (Resource Lifetime)}}}
  & CWE-119 (Memory Bounds)    & 6 & 188 & 29.3 \\
  & CWE-362 (Race Condition)   & 3 &  85 & 13.2 \\
  & CWE-404 (Resource Release) & 1 &  51 &  7.9 \\
  & CWE-665 (Initialization)   & 2 &  50 &  7.8 \\
  & CWE-667 (Locking)          & 2 &  50 &  7.8 \\
  & CWE-704 (Type Conversion)  & 3 &  50 &  7.8 \\
\midrule
\makecell[l]{CWE-710\textit{ (Coding Standards)}}
  & ---                        & 1 &  66 & 10.3 \\
\makecell[l]{CWE-682\textit{ (Incorrect Calc.)}}
  & ---                        & 4 &  52 &  8.1 \\
\makecell[l]{CWE-707\textit{ (Neutralization)}}
  & CWE-20 (Input Validation)  & 4 &  50 &  7.8 \\
\midrule
\multicolumn{2}{l}{\textbf{Total (4 Pillars, 7 Classes, 26 CWEs)}}
  & \textbf{---} & \textbf{642} & \textbf{100} \\
\bottomrule
\end{tabularx}
\end{table}

\begin{table}[!ht]
\caption{{\small Mechanism-level taxonomy of the sampled eBPF vulnerabilities. 
Count denotes the number of sampled instances assigned to each category or subcategory.}}
\label{tab:taxonomy}
\centering
\small
\renewcommand{\arraystretch}{0.8}
\setlength{\tabcolsep}{4pt}
\begin{tabularx}{0.9\linewidth}{llXl}
\toprule
\textbf{Cat.} & \textbf{Scope} & \textbf{Subcategory} & \textbf{Count} \\
\midrule

\multirow{5}{*}{\textbf{A}}
 & \multirow{5}{*}{Verifier (114, 17.8\%)}
 & A1: Scalar / range reasoning defects & 30 \\
 & & A2: Pointer / type reasoning defects & 29 \\
 & & A3: State merge / pruning / patching defects & 21 \\
 & & A4: Helper / kfunc behavior modeling defects & 14 \\
 & & A5: Verifier implementation bugs & 20 \\

\midrule

\multirow{4}{*}{\textbf{B}}
 & \multirow{4}{*}{JIT compiler (54, 8.4\%)}
 & B1: Instruction selection / encoding errors & 20 \\
 & & B2: Register / stack / calling convention errors & 11 \\
 & & B3: Architecture / ABI semantic errors & 12 \\
 & & B4: JIT memory management errors & 11 \\

\midrule

\multirow{6}{*}{\textbf{C}}
 & \multirow{6}{*}{Runtime (420, 65.4\%)}
 & C1: Helper implementation defects & 58 \\
 & & C2: Map / storage implementation defects & 28 \\
 & & C3: Kernel object lifecycle / refcount defects & 80 \\
 & & C4: Concurrency / synchronization defects & 135 \\
 & & C5: BPF execution infrastructure defects & 19 \\
 & & C6: Kernel subsystem flaws exposed via eBPF & 100 \\

\midrule

\textbf{D} & Userspace (43, 6.7\%) & (no subdivision) & 43 \\

\midrule

\multirow{3}{*}{\textbf{E}}
 & \multirow{3}{*}{Cross-component (11, 1.7\%)}
 & E1: Verifier--Runtime mismatch & 7 \\
 & & E2: JIT--Interpreter mismatch & 1 \\
 & & E3: Userspace--Kernel ABI mismatch & 3 \\

\midrule
\multicolumn{3}{l}{\textbf{Total}} & \textbf{642} \\
\bottomrule
\end{tabularx}
\end{table}

Table~\ref{tab:taxonomy} presents the statistical results of the mechanism-level taxonomy used in this study. 
Under this taxonomy, 
we organize sampled vulnerabilities according to the system-level failure modes that underlie them, 
rather than only their surface code-level defect forms. 
The structural basis for this decomposition derives from the trust model formalized by Huang~et~al.~\cite{huang2025sok}, 
together with prior eBPF security research on the verifier~\cite{sun2024validating,gershuni2019simple}, 
the JIT compiler~\cite{nelson2020specification}, 
the runtime~\cite{hung2024brf}, 
and cross-component semantic mismatches~\cite{peng2024toss,jia2025rex}. 
By consolidating these perspectives into a unified classification framework, 
we categorize vulnerabilities according to a consistent set of mechanism-level fault modes 
spanning the major stages and components of the eBPF execution pipeline.

Unlike traditional root-cause classifications that focus primarily on code-level defect types, 
our taxonomy emphasizes the system execution stage and semantic component at which a security guarantee fails. 
In other words, vulnerabilities are classified according to \emph{which system mechanism fails to enforce the intended security property}. 
To our knowledge, 
this is among the first efforts to consolidate security defects across eBPF subsystem components 
into a unified mechanism-level taxonomy grounded in a systematically collected vulnerability dataset.
In the following, 
we first use representative case studies to illustrate how dominant mechanism-level failures manifest in practice, 
and then quantify their overall concentration in the sampled vulnerability landscape.

\subsubsection{Representative Case Studies} \label{sec:rq2:cases}
We next examine a set of representative cases that 
show how the dominant mechanism-level failures arise in practice. 
Each case is analyzed around two perspectives:
(i)~at which stage and why the failure occurs;
(ii)~which security invariant is violated.

\noindent
\textbf{Case~\blackcircleone{1}: Pointer / type reasoning defects [CVE-2022-23222].}\label{para:case1}
This vulnerability arises from incomplete semantic constraints 
in the BPF verifier's handling of \texttt{*\_OR\_NULL} pointer types. 
Since Linux~5.8, 
the verifier has introduced several nullable pointer types 
(e.g., \texttt{PTR\_TO\_MEM\_OR\_NULL}) to represent helper return values that may be \texttt{NULL}. 
However, 
it did not prohibit pointer arithmetic on these nullable pointers before a \texttt{NULL} check, 
allowing an attacker to create a divergence 
between the verifier's abstract type state and the actual runtime value.

\begin{figure}[!ht]
\tcbset{colback=gray!10,colframe=gray!40,boxrule=0.3pt}
\begin{tcolorbox}
\ttfamily\smaller
\linespread{0.95}
\begin{verbatim}
static int adjust_ptr_min_max_vals(...) {
    ...
    switch (ptr_reg->type) {
    case PTR_TO_MAP_VALUE_OR_NULL:       /* blocked */
        ...
        return -EACCES;
    ...
    case PTR_TO_SOCKET_OR_NULL:          /* blocked */
    case PTR_TO_SOCK_COMMON_OR_NULL:     /* blocked */
    case PTR_TO_TCP_SOCK_OR_NULL:        /* blocked */
    case PTR_TO_XDP_SOCK:
        ...
        return -EACCES;
    default:
        break;                           /* PTR_TO_MEM_OR_NULL not listed -- arithmetic allowed */
    }
    dst_reg->type = ptr_reg->type;       /* Arithmetic proceeds; dst inherits ptr type. */
    dst_reg->id   = ptr_reg->id;
    ...
}
\end{verbatim}
\end{tcolorbox}
\caption{[Case-1]Missing pointer-arithmetic restriction for \texttt{PTR\_TO\_MEM\_OR\_NULL} in \texttt{adjust\_ptr\_min\_max\_vals()} (\texttt{kernel/bpf/verifier.c}, before fix). Because this nullable pointer type is not included in the verifier's rejection logic, pointer arithmetic is silently permitted before the \texttt{NULL} check.}
\label{fig:cve-2022-23222}
\end{figure}

As shown in Figure~\ref{fig:cve-2022-23222}, 
\texttt{adjust\_ptr\_min\_max\_vals()} uses a \texttt{switch} statement to 
reject arithmetic on known \texttt{*\_OR\_NULL} pointer types. 
However, 
\texttt{PTR\_TO\_MEM\_OR\_NULL}, introduced in Linux~5.8 as the return type of \texttt{bpf\_ringbuf\_re} \texttt{serve()}, 
is absent from this list and therefore falls through 
to \texttt{default:\ break}, silently permitting pointer arithmetic.
Because arithmetic is permitted, an attacker can construct the following path: 
perform \texttt{r1 = r0;\ r1 += 1} on the return value \texttt{r0} of \texttt{bpf\_ringbuf\_reserve()} 
(typed \texttt{PTR\_TO\_MEM\_OR\_NULL}). 
In the subsequent \texttt{NULL} branch, when \texttt{r0 == NULL}, 
the verifier propagates the conclusion that \texttt{r0} is \texttt{NULL} to \texttt{r1}, 
and therefore concludes that \texttt{r1} is also~0. 
At runtime, however, \texttt{r1} is actually~1 because the arithmetic operation has already been executed. 
This mismatch between verifier state and runtime value can then be exploited for arbitrary read/write.

\noindent
\textbf{Violated invariant.}
Nullable pointer types must not participate in arithmetic that 
changes their abstract value before the \texttt{NULL} check is completed. 
Moreover, 
\texttt{NULL}-branch reasoning must propagate consistently to all aliased registers, 
including those affected by prior arithmetic transformations.

\noindent
\textbf{Conclusion.}
This case shows that pointer/type reasoning defects arise 
from incomplete constraints in the verifier's abstract type system. 
When new pointer types are introduced without corresponding updates to arithmetic restrictions, 
the verifier may allow divergence between abstract state and runtime value, 
creating an exploitable gap. 
Because this failure occurs at the verifier's type-rule level, 
it cannot be reliably mitigated by runtime checks and must instead be addressed through sound verifier semantics.

\noindent
\textbf{Case~\blackcircleone{2}: Architecture/ABI convention defects (JIT) [CVE-2025-22048].}\label{para:case2}
This vulnerability illustrates how JIT-generated machine code 
can diverge semantically from the bytecode semantics already approved by the verifier.
In the BPF security model, 
the verifier establishes safety guarantees at the bytecode level by checking 
whether a program satisfies the required semantic and safety constraints before execution, 
whereas the JIT is responsible for faithfully translating that verified bytecode into target-architecture machine code. 
This design implicitly assumes that the machine code generated by the JIT preserves the semantics already approved by the verifier. 
When the JIT introduces extra semantics not present in the original bytecode, 
such as unintended register overwrites, 
it breaks this assumption and creates a security gap that the verifier cannot detect.

\begin{figure}[!ht]
\tcbset{colback=gray!10,colframe=gray!40,boxrule=0.3pt}
\begin{tcolorbox}
\ttfamily\smaller
\linespread{1}
\begin{verbatim}
static int build_insn(const struct bpf_insn *insn, struct jit_ctx *ctx, bool extra_pass) {
    ...
    case BPF_JMP | BPF_CALL:
        ...
        move_addr(ctx, t1, func_addr);
        emit_insn(ctx, jirl, LOONGARCH_GPR_RA, t1, 0);
        move_reg(ctx, regmap[BPF_REG_0], LOONGARCH_GPR_A0);
         /*        ^^^ a0 -> a5 unconditionally.
         * Correct for native calls (helper ABI uses a0),
         * WRONG for bpf2bpf calls: subprog already wrote
         * zero-extended result into a5 directly;
         * a0 holds sign-extended copy -> corrupts result. */
        break;
    ...
}

if (insn->src_reg != BPF_PSEUDO_CALL)   /* Fix (commit 60f3caff1492): only for native calls */
    move_reg(ctx, regmap[BPF_REG_0], LOONGARCH_GPR_A0);
\end{verbatim}
\end{tcolorbox}
\caption{[Case-2]Unconditional \texttt{a0}$\to$\texttt{BPF\_REG\_0} move in \texttt{build\_insn()} (\texttt{arch/loongarch/net/bpf\_jit.c}, before fix), 
causing the JIT to overwrite the correct return register state after bpf2bpf calls.}
\label{fig:cve-2025-22048}
\end{figure}

On the LoongArch architecture, 
the BPF return-value register \texttt{BPF\_REG\_0} is mapped to \texttt{a5}, 
whereas the native LoongArch ABI uses \texttt{a0} as the function return register. 
A prior commit (73c359d1d356) introduced a post-call \texttt{a0}$\to$\texttt{a5} move 
to accommodate the native ABI's sign-extension requirements. 
As shown in Figure~\ref{fig:cve-2025-22048}, 
\texttt{build\_insn()} unconditionally copies \texttt{LOONGARCH\_GPR\_A0} 
into \texttt{regmap[BPF\_REG\_0]} (\texttt{a5}) after every function call. 
For native helper calls, 
this behavior is correct because the helper returns its value in \texttt{a0}, 
which must then be forwarded to \texttt{a5}. 
However, 
for bpf2bpf subprogram calls (\texttt{BPF\_PSEUDO\_CALL}), 
the callee has already written the zero-extended return value directly into \texttt{a5}. 
The extra JIT-emitted \texttt{move} instruction 
therefore overwrites the correct value in \texttt{a5} 
with the sign-extended value in \texttt{a0}—an operation that does not exist in the original BPF bytecode and 
is introduced solely by the JIT translation.
The issue was triggered by the verifier test \texttt{calls:\ div\ by\ 0\ in\ subprog}, 
which caused a panic. 
The subprogram first wrote the correct memory address into \texttt{a5}, 
but after the call returned, 
the JIT-emitted \texttt{move\_reg} instruction 
overwrote \texttt{a5} with the sign-extended value from \texttt{a0}, 
then
a subsequent \texttt{ld.bu} instruction accessed an incorrect memory address.

\noindent
\textbf{Violated invariant.}
The machine code generated by the JIT must faithfully 
preserve the bytecode semantics already validated by the verifier. 
Any extra instruction introduced during JIT translation 
must not alter the semantic state of BPF registers beyond what is expressed in the original bytecode.

\noindent
\textbf{Conclusion.}
This case reveals the trust-boundary risk 
between the verifier and the JIT in the BPF security model. 
The verifier establishes safety guarantees at the bytecode level, 
but the JIT may introduce semantic changes absent from the verified program, 
creating a security gap beyond the verifier's reach. 
Such machine-code-level semantic divergence is architecture-specific (e.g., LoongArch) and 
cannot be eliminated simply by strengthening bytecode-level verification. 
The underlying issue is that the JIT, 
as a trusted translation layer, 
lacks a systematic correctness-verification mechanism.

\noindent
\textbf{Case~\blackcircleone{3}: Kernel object lifecycle/reference-count defects [CVE-2022-50219].}\label{para:case3}
This vulnerability arises from an object-lifecycle management defect in the link-detach error-recovery path of the cgroup BPF subsystem. 
In this subsystem, BPF programs can be attached to cgroups through \texttt{bpf\_link}, and detaching a link requires updating the effective-programs array of the target cgroup and all its descendants. 
As shown in Figure~\ref{fig:cve-2022-50219}, when \texttt{update\_effective\_progs()} fails due to a memory allocation error, the error-recovery path restores the about-to-be-freed \texttt{link} pointer back into the list, thereby creating a dangling reference.

\begin{figure}[!ht]
\tcbset{colback=gray!10,colframe=gray!40,boxrule=0.3pt}
\begin{tcolorbox}
\ttfamily\smaller
\linespread{0.95}
\begin{verbatim}
int __cgroup_bpf_detach(struct cgroup *cgrp, ...) {
    ...
    int err;
    pl->prog = NULL;
    pl->link = NULL;
    err = update_effective_progs(cgrp, atype);
    if (err)
        goto cleanup;
    list_del(&pl->node);            /* normal path: delete and free */
    ...
    return 0;
cleanup:
    /* restore back prog or link */
    pl->prog = old_prog;            /* <- will dangle */
    pl->link = link;                /* <- will dangle */
    return err;
}
/* After return, bpf_link_free() frees 'link'.
 * The cgroup prog list now holds a dangling pointer;
 * any subsequent compute_effective_progs() call
 * dereferences the freed object -> KASAN UAF.        */
\end{verbatim}
\end{tcolorbox}
\caption{[Case-3]Vulnerable error-recovery path in \texttt{\_\_cgroup\_bpf\_detach()} (\texttt{kernel/bpf/cgroup.c}, before fix), where the cleanup logic restores a link pointer that is freed immediately after return.}
\label{fig:cve-2022-50219}
\end{figure}

This defect can be triggered even without concurrency. 
Within \texttt{update\_effective\_progs()}, a single thread can force the allocation to fail via fault injection. 
After the error path executes, the \texttt{link} object is freed, and any subsequent operation that recomputes the effective-programs array dereferences an already-freed object.

\vspace{2pt}
\noindent
\textbf{Violated invariant.}
Object references stored in kernel data structures must remain valid throughout their reachable lifetime. 
In particular, error-recovery paths must not restore references to objects whose lifecycle is about to end.
The root cause is an object-lifecycle boundary violation: \texttt{bpf\_link\_free()} releases the object immediately after the function returns, 
but the error path has already written that same address back into the list. 
This creates a classic error-path lifecycle mismatch, in which cleanup logic reconstructs a reference that is no longer safe to retain.

\noindent
\textbf{Conclusion.}
This case shows that object-lifecycle defects arise from inconsistent lifetime assumptions between normal paths and error-recovery paths. 
BPF objects span multiple management layers, including user references, cgroup hierarchy state, and effective-program arrays, and their lifetime is determined jointly by multiple release paths. 
When an error path attempts to roll back a partial operation without respecting these lifetime relationships, dangling references and use-after-free conditions can result.

\noindent
\textbf{Case~\blackcircleone{4}: Concurrency/synchronization defects [CVE-2023-0160].}\label{para:case4}
This vulnerability is a runtime deadlock in the BPF sockmap/sockhash subsystem. 
It arises because the locking discipline in \texttt{sock\_hash\_delete\_elem()} 
does not account for the fact that eBPF-triggered map operations may execute in hardirq context. 
As shown in Figure~\ref{fig:cve-2023-0160}, the function uses \texttt{raw\_spin\_lock\_bh()} 
to acquire the bucket lock, which disables softirqs but leaves hardirqs enabled. 
As a result, if a hardirq interrupts the lock-holding path and attempts to acquire the same lock, 
or another lock that participates in a circular dependency (e.g., \texttt{rq->\_\_lock}), 
a deadlock can occur.

\begin{figure}[!ht]
\tcbset{colback=gray!10,colframe=gray!40,boxrule=0.3pt}
\begin{tcolorbox}
\ttfamily\smaller
\linespread{0.95}
\begin{verbatim}
static long sock_hash_delete_elem(struct bpf_map *map, void *key) {
    ...
    hash   = sock_hash_bucket_hash(key, key_size);
    bucket = sock_hash_select_bucket(htab, hash);
    raw_spin_lock_bh(&bucket->lock);
    /*           ^^^ softirq disabled, hardirq still enabled.
     * eBPF programs (XDP/tc) can invoke this function in
     * hardirq context -> deadlock if interrupt arrives while
     * this lock is held.                                    */
    elem = sock_hash_lookup_elem_raw(&bucket->head, hash,
                                     key, key_size);
    if (elem) {
        hlist_del_rcu(&elem->node);
        sock_map_unref(elem->sk, elem);
        sock_hash_free_elem(htab, elem);
        ret = 0;
    }
    raw_spin_unlock_bh(&bucket->lock);
    return ret;
}

    /* Fix (commit ed17aa92dc56): replace _bh with _irqsave */
    raw_spin_lock_irqsave(&bucket->lock, flags);
    ...
    raw_spin_unlock_irqrestore(&bucket->lock, flags);
\end{verbatim}
\end{tcolorbox}
\caption{[Case-4]Vulnerable locking in \texttt{sock\_hash\_delete\_elem()} (\texttt{net/core/sock\_map.c}, before fix). 
Using \texttt{\_bh} disables softirqs but leaves hardirqs enabled, allowing an interrupt-time reentry that can deadlock on the same lock.}
\label{fig:cve-2023-0160}
\end{figure}

\begin{figure}[!ht]
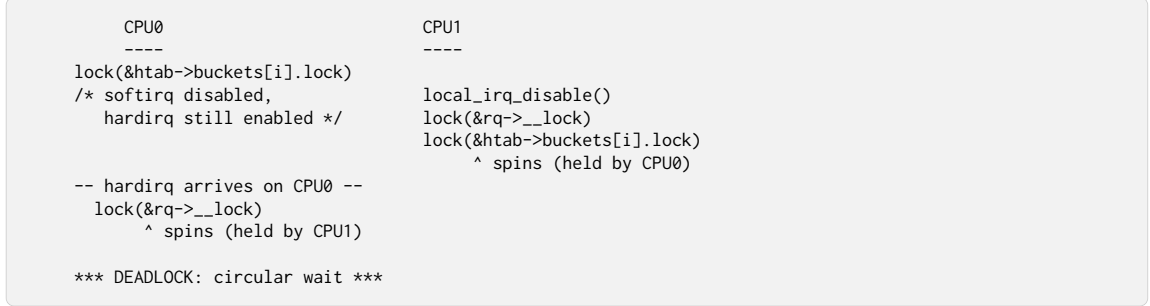

\tcbset{colback=gray!10,colframe=gray!40,boxrule=0.3pt}
\begin{tcolorbox}
\ttfamily\smaller
\linespread{0.95}
\begin{verbatim}
        CPU0                          CPU1
        ----                          ----
   lock(&htab->buckets[i].lock)
   /* softirq disabled,               local_irq_disable()
      hardirq still enabled */        lock(&rq->__lock)
                                      lock(&htab->buckets[i].lock)
                                           ^ spins (held by CPU0)
   -- hardirq arrives on CPU0 --
     lock(&rq->__lock)
          ^ spins (held by CPU1)

   *** DEADLOCK: circular wait ***
\end{verbatim}
\end{tcolorbox}
\caption{[Case-4]Lockdep deadlock scenario for CVE-2023-0160. 
A hardirq on CPU0 interrupts the bucket-lock critical section, while CPU1 already holds \texttt{rq->\_\_lock}, producing a circular wait between the two locks.}
\label{fig:cve-2023-0160-deadlock}
\end{figure}

\noindent
\textbf{Violated invariant.}
The lock-acquisition order for shared data structures must maintain a consistent partial order across all execution contexts, 
including process, softirq, and hardirq, and must not create circular dependencies.
The vulnerability was reproduced via lockdep on Linux v5.15.25 and v5.19. 
As illustrated in Figure~\ref{fig:cve-2023-0160-deadlock}, 
CPU0 first acquires the bucket lock, after which a hardirq arrives and attempts to acquire \texttt{rq->\_\_lock}; 
at the same time, CPU1 already holds \texttt{rq->\_\_lock} and is waiting for the bucket lock. 
This interleaving creates a circular wait and therefore a deadlock.
The core issue is that eBPF programmability allows map operations to execute in hardirq context, 
whereas the existing lock protection only accounts for softirq-level concurrency. 
Because the verifier does not model runtime lock ordering across execution contexts, 
this defect cannot be detected at verification time.

\noindent
\textbf{Conclusion.}
This case shows that concurrency/synchronization defects arise 
when eBPF extends kernel operations into execution contexts that existing locking assumptions do not fully cover. 
Traditional kernel reasoning may assume that a given path executes only in process or softirq context, 
but under eBPF the same operation may also be triggered in hardirq and other contexts. 
As a result, concurrency safety in eBPF must be understood not only in terms of individual locks, 
but also in terms of whether lock ordering remains sound across all reachable execution contexts.

\noindent
\textbf{Case~\blackcircleone{5}: Kernel subsystem flaws exposed via eBPF [CVE-2024-26611].}\label{para:case5}
This vulnerability arises because the AF\_XDP (XSK) subsystem's zero-copy buffer-management logic 
does not account for a packet-shrink path exposed through a BPF helper. 
Linux~v6.4 introduced multi-buffer zero-copy receive support, 
allowing XDP programs to process packets that span multiple buffers. 
The helper \texttt{bpf\_xdp\_adjust\_tail()} allows an XDP program to shrink the packet length at runtime; 
when this operation completely removes the last fragment, the corresponding backing memory must be released correctly.
As shown in Figure~\ref{fig:cve-2024-26611}, 
the original release path only handles page-backed memory.
In zero-copy mode, 
however, 
buffers are managed by the XSK buffer pool rather than the page allocator. 
As a result, \texttt{skb\_frag\_page()} returns \texttt{NULL}, 
and the subsequent call to \texttt{page\_address(NULL)} triggers a kernel NULL-pointer dereference. 
The key point is that the subsystem release logic is correct for traditional page-backed buffers, 
but incomplete for the helper-reachable execution path introduced by zero-copy XSK buffers.

\begin{figure}[!ht]
\tcbset{colback=gray!10,colframe=gray!40,boxrule=0.3pt}
\begin{tcolorbox}
\ttfamily\smaller
\linespread{0.95}
\begin{verbatim}
static int bpf_xdp_frags_shrink_tail(struct xdp_buff *xdp, int offset) {
    ...
    if (skb_frag_size(frag) == shrink) {
        struct page *page = skb_frag_page(frag);
        /* Assumes page-backed memory -- correct for
           MEM_TYPE_PAGE_SHARED / MEM_TYPE_PAGE_ORDER0.
           But for MEM_TYPE_XSK_BUFF_POOL (zero-copy),
           there is NO page; skb_frag_page() yields NULL. */
        __xdp_return(page_address(page), &xdp->rxq->mem,
                     false, NULL);
        /*            ^^^^^^^^^^^^^^^^
           NULL ptr dereference when mem.type == XSK_BUFF_POOL */
        n_frags_free++;
    } ...
}
\end{verbatim}
\end{tcolorbox}
\caption{[Case-5]Vulnerable shrink path in \texttt{bpf\_xdp\_frags\_shrink\_tail()} (\texttt{net/core/filter.c}, before fix), 
where the release logic assumes page-backed memory and fails for zero-copy XSK buffer-pool memory.}
\label{fig:cve-2024-26611}
\end{figure}

\noindent
\textbf{Violated invariant.}
Kernel subsystems must preserve resource and state consistency 
regardless of how an execution path is reached. 
In particular, the buffer-release path must correctly distinguish 
among all supported memory backend types.
The issue is neither a verifier bug nor a logic error in the helper itself. 
Rather, the XSK subsystem's buffer-management logic 
does not fully cover execution paths that become reachable through BPF helpers. 
eBPF programmability transforms buffer operations that were previously triggered only 
by subsystem-internal logic into user-triggerable execution paths.

\noindent
\textbf{Conclusion.}
This case shows how eBPF can amplify the attack surface of existing kernel subsystems. 
By exposing internal subsystem operations through programmable helper-triggered paths, 
eBPF can make latent subsystem assumptions externally reachable and therefore security-relevant. 
The underlying mechanism is not a failure of verification or bytecode semantics, 
but a mismatch between subsystem implementation assumptions and 
the broader set of execution paths that eBPF makes reachable.

\noindent
\textbf{Mechanism-Level Insights.}
These five cases illustrate the dominant mechanism-level failure modes underlying real-world eBPF vulnerabilities and 
show how they arise across the major stages of the eBPF execution pipeline:

\begin{itemize}
  \item \textbf{Case~1: Type-reasoning failure at the verification stage.}
    The verifier's abstract type system did not include a newly introduced pointer type in its arithmetic restrictions, creating an exploitable divergence between verifier state and runtime values.
  \item \textbf{Case~2: Semantic divergence at the JIT compilation stage.}
    The JIT translation process introduced a register-overwrite operation absent from the bytecode, causing machine-code semantics to diverge from the verifier-approved bytecode semantics and creating a security gap beyond the verifier's reach.
  \item \textbf{Case~3: Runtime object-lifecycle management failure.}
    The root cause lies in inconsistent lifetime assumptions between error-recovery paths and normal paths; the error path restores a reference to an object that is about to be freed.
  \item \textbf{Case~4: Runtime concurrency consistency failure.}
    eBPF's multi-context programmability allows the same operation to execute under different interrupt contexts, breaking concurrency assumptions in the existing kernel design and rendering softirq-only lock protection ineffective once the path becomes reachable in hardirq context.
  \item \textbf{Case~5: Structural exposure risk of cross-subsystem execution paths.}
    eBPF transforms subsystem-internal execution paths into user-triggerable entry points, exposing otherwise unreachable internal flaws.
\end{itemize}

Viewed together, these cases answer RQ2 by showing that 
dominant real-world eBPF vulnerabilities are underlain not by many unrelated low-level defect forms, 
but by a limited set of recurring system-level failures. 
These failures span verifier reasoning, JIT semantic preservation, runtime lifecycle management, 
concurrency control, and cross-subsystem reachability. 
Compared with CWE-based statistics alone, 
this mechanism-level view more directly explains where dominant vulnerabilities arise and why they recur, 
providing a stronger basis for the architectural analysis in RQ3 and the technique assessment in RQ4.

\subsubsection{Mechanism-Level Concentration Analysis}

\begin{figure}[!ht]
\centering
\includegraphics[width=\linewidth]{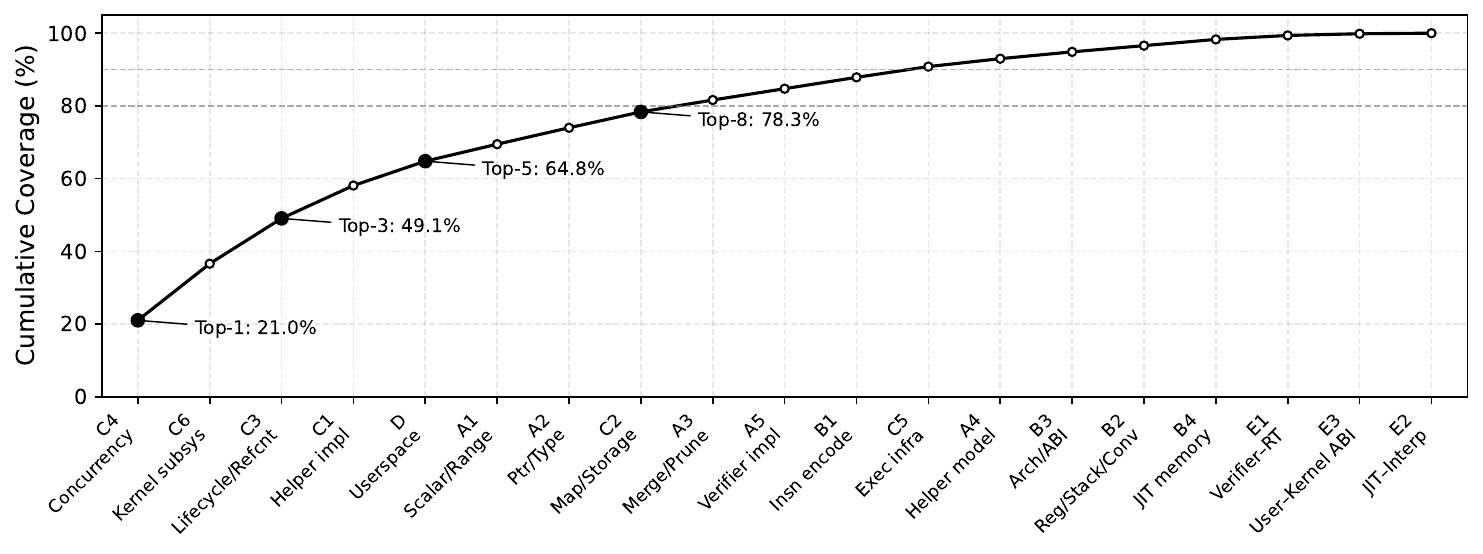}
\caption{Cumulative coverage of mechanism subtypes in the sampled vulnerability taxonomy. 
The figure shows that a relatively small number of mechanism-level failure modes account for most sampled vulnerabilities.}
\label{fig:Mechanism_Cumulative}
\end{figure}

To further answer RQ2, 
we quantify how sampled vulnerabilities are distributed across the mechanism subcategories identified above. 
Figure~\ref{fig:Mechanism_Cumulative} shows the cumulative coverage of these subtypes. 
The results reveal a clear concentration pattern at the mechanism level. 
In particular, the most frequent subtype, C4 (concurrency/synchronization defects), accounts for 21.0\% of the sample. 
When the top three subtypes (C4, C6, and C3) are considered together, their cumulative coverage reaches 49.1\%. 
Expanding to the top five subtypes increases this to 64.8\%, 
while the top eight subtypes cover 78.3\% of the sample.

Overall, 
the mechanism-level landscape is strongly skewed: 
a small number of subtypes account for the majority of sampled cases. 
This concentration is driven primarily by runtime-related failures, 
especially concurrency/synchronization defects, 
kernel subsystem flaws exposed via eBPF, 
and kernel object lifecycle/reference-count defects. 
The cumulative curve rises rapidly before flattening, 
indicating that the dominant portion of the sampled vulnerability landscape 
is explained by a relatively small set of recurring mechanism-level failure modes.
These results strengthen the answer to RQ2. 
The sampled analysis of dominant strata suggests that 
real-world eBPF vulnerabilities are underlain by a limited set of recurring system-level failure modes, 
rather than a broad collection of unrelated defect types. 
This concentration provides the basis for the subsequent architectural analysis in RQ3 
and the technique assessment in RQ4.

\begin{tcolorbox}[colback=gray!15,colframe=gray!60,boxrule=0.8pt,arc=4pt]
\textbf{TA 3. The sampled mechanism-level analysis suggests that 
the dominant observed portion captured by the sampled strata of the eBPF vulnerability landscape is associated
with a limited set of recurring system-level failure modes, rather than a broad collection of unrelated defect types. 
In particular, these failures concentrate in runtime execution, concurrency, object lifecycle management, 
and semantic inconsistencies across trusted stages. 
These findings answer RQ2 and provide the basis for architectural analysis in RQ3.}
\end{tcolorbox}

\subsection{RQ3: How are the dominant failure mechanisms distributed across major eBPF components and execution stages?}\label{sec:rq3}

The results show that the dominant mechanism-level failures identified in RQ2 
are unevenly distributed across the eBPF architecture. 
Runtime accounts for the largest share of observed vulnerabilities,
while the Verifier and JIT contribute smaller but structurally distinct classes of failure. 
The following analyses explain this distribution through the full-dataset component breakdown, 
the sampled mechanism-level results, and the execution-stage roles defined by the eBPF trust model.

\subsubsection{Empirical Distribution of Vulnerabilities Across Components} \label{sec:rq3:comp-dist}

\begin{figure}[!ht]
\centering
\includegraphics[width=0.8\linewidth]{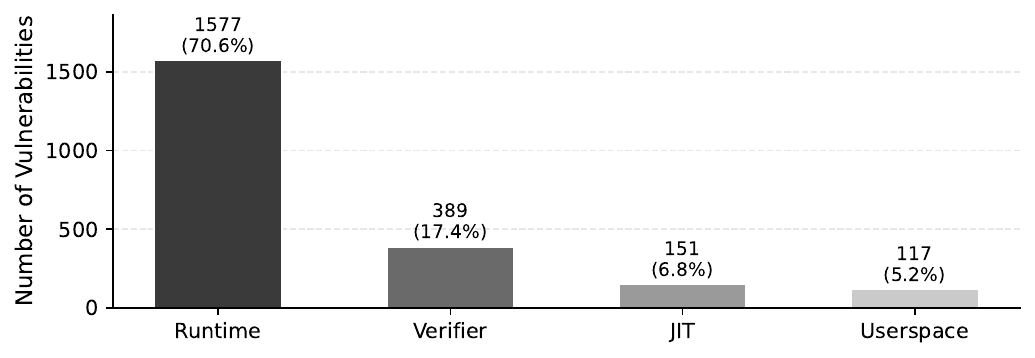}
\caption{Distribution of vulnerability instances across major eBPF components under a single component assignment per instance. Runtime accounts for the largest share of observed vulnerabilities.}
\label{fig:component_distribution_bw}
\end{figure}

Figure~\ref{fig:component_distribution_bw} shows the distribution of vulnerability instances across major eBPF components 
under a single component assignment per instance. 
Unlike Table~\ref{tab:cwe-level-dist-component}, which reports level-specific counts by assigned CWE abstraction level, Figure~\ref{fig:component_distribution_bw} summarizes the component distribution at the instance level.
Overall, 
the Runtime component accounts for 1{,}577 instances (70.6\%), 
while the Verifier accounts for 389 (17.4\%), and JIT and Userspace contribute 151 (6.8\%) and 117 (5.2\%), respectively. 
This result shows that real-world eBPF vulnerabilities are not uniformly distributed across components, 
but are strongly concentrated in Runtime-related paths associated with program execution and system interaction.

This concentration is also consistent with the sampled mechanism-level analysis in Section~\ref{sec:rq2}. 
Among the 642 sampled instances, Runtime-related cases account for 420 (65.4\%), 
while the Verifier, JIT, and Userspace account for 114, 54, and 43 instances, respectively. 
Viewed together, 
the full-dataset statistics and the sampled mechanism-level results indicate that 
Runtime constitutes the dominant empirical exposure area in the observed vulnerability landscape.
At the same time, component-level frequency alone is not sufficient 
to characterize the architectural role or security significance of each component. 
To answer RQ3 more completely, 
we therefore interpret this concentration together with the component-specific weakness composition, 
the execution-stage responsibilities defined by the eBPF trust model, and the representative mechanism-level cases from RQ2.

\subsubsection{Component-Level Risk Interpretation via Execution Flow and Trust Model}
\label{sec:rq3:trust-model}

\begin{figure}[!ht]
\centering
\includegraphics[width=0.9\linewidth]{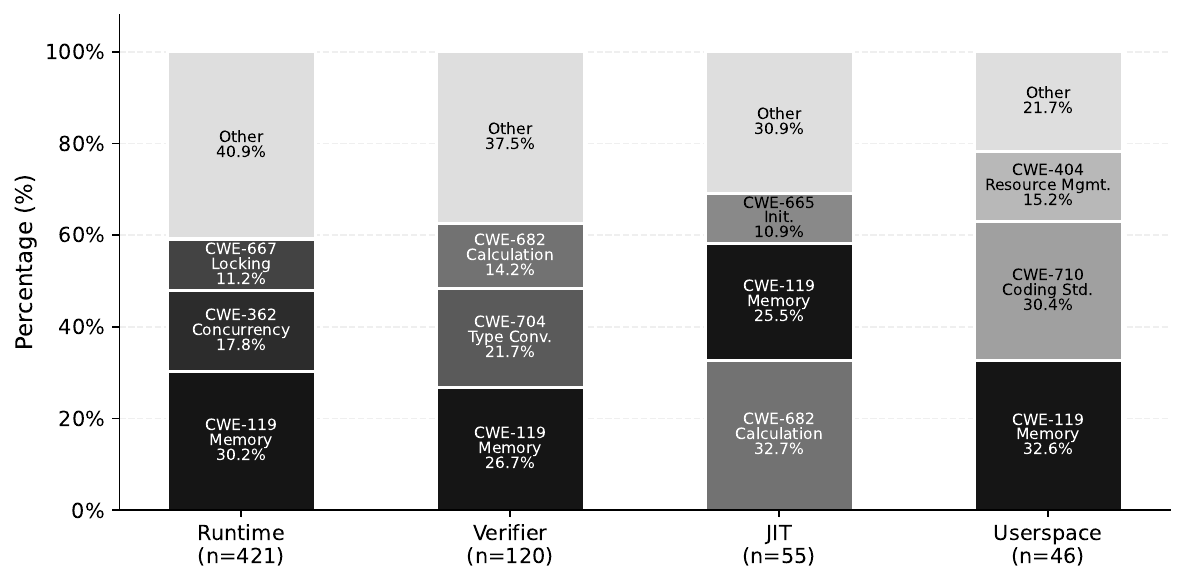}
\caption{Top-3 CWE categories within each major eBPF component, normalized by within-component percentage. The results show that different components exhibit distinct weakness compositions rather than a uniform vulnerability profile.}
\label{fig:component_cwe_composition}
\end{figure}

Component-level frequency alone is not sufficient 
to explain the architectural significance of the observed vulnerability distribution.
To answer RQ3 more completely, 
we interpret the component-wise results together 
with the execution flow and trust relationships in the eBPF security model.
As illustrated in Figure~\ref{fig:work_flow_and_trust_model}, 
eBPF programs enter the kernel from user space as untrusted input, 
are checked by the Verifier, translated by the JIT or interpreter, 
and then execute in the Runtime while interacting with maps, helpers, and other kernel subsystems.
These stages play different security roles: 
Runtime is the primary execution exposure surface, 
the Verifier is the security-enforcement boundary, 
and the JIT is the semantic-preservation boundary.
Accordingly, 
the same problem type can have different security significance depending on 
where it occurs in the execution flow.
Figure~\ref{fig:component_cwe_composition} further shows that 
the dominant weakness composition differs substantially across components, 
indicating that the major eBPF components correspond 
to distinct risk-bearing surfaces in the architecture.

\vspace{2pt}
\noindent
\textbf{Runtime.}
Runtime, the component with the highest vulnerability concentration, 
is dominated by memory-related weaknesses (CWE-119, 30.2\%), concurrency defects (CWE-362, 17.8\%), 
and locking-related issues (CWE-667, 11.2\%).
This composition is consistent with Runtime’s role in the eBPF security model.
Unlike the Verifier and JIT, 
Runtime directly handles concrete execution-time interactions 
with maps, helpers, object lifecycles, and other kernel subsystems.
Accordingly, its dominant risks arise not from abstract semantic judgment, 
but from execution-time failures involving memory access, concurrent access, resource coordination, 
and subsystem state consistency.
The concentration observed here therefore reflects the fact that 
Runtime bears the broadest and most operationally exposed attack surface in the eBPF architecture.

\vspace{2pt}
\noindent
\textbf{Verifier.}
The Verifier exhibits a markedly different weakness profile.
As shown in Figure~\ref{fig:component_cwe_composition}, 
its dominant categories include memory-related issues (CWE-119, 26.7\%), type-conversion errors (CWE-704, 21.7\%), 
and computation errors (CWE-682, 14.2\%).
This pattern is consistent with the Verifier’s role in screening untrusted bytecode before execution.
Its core responsibility is to determine whether untrusted bytecode 
satisfies the semantic and safety constraints required to enter kernel execution.
As a result, 
its dominant risks are concentrated in the soundness of abstract reasoning itself: 
whether types are modeled correctly, 
whether numeric constraints are propagated soundly, 
and whether corner cases are handled consistently.
Failures at this stage therefore represent unsound safety judgment 
rather than ordinary runtime execution defects.

\vspace{2pt}
\noindent
\textbf{JIT.}
Although the JIT contributes fewer vulnerabilities overall, 
its internal composition is also structurally distinct.
Its dominant categories are computation errors (CWE-682, 32.7\%), 
memory-related issues (CWE-119, 25.5\%), and initialization errors (CWE-665, 10.9\%).
Positioned between bytecode verification and machine-level execution, 
the JIT serves as the trusted translation layer from verified bytecode to machine code.
Accordingly, the significance of its risk lies less in raw frequency than in architectural role: 
errors in this stage can cause the machine code 
to diverge from the bytecode semantics already approved by the Verifier.
The weakness composition of the JIT therefore reflects the risk of translation-side semantic divergence, 
especially when architecture-specific behavior fails to preserve verifier-approved program semantics.

\vspace{2pt}
\noindent
\textbf{Userspace.}
Userspace shows yet another profile, 
with memory-related issues (CWE-119, 32.6\%), 
coding-standard and implementation-quality issues (CWE-710, 30.4\%), 
and resource-management defects (CWE-404, 15.2\%) as its dominant categories.
Compared with Runtime, Verifier, and JIT, 
this composition is more characteristic of engineering robustness issues in tools and supporting ecosystem components.
Although such issues remain relevant to the broader eBPF ecosystem, 
their architectural security role is less central than that of the core kernel-side stages.

These component-specific weakness profiles show that 
the observed vulnerability concentration is not simply an uneven distribution of counts. 
Instead, 
it reflects a deeper architectural differentiation across the eBPF execution flow, 
which helps explain why the dominant failure mechanisms identified in RQ2 
are distributed unevenly across components and stages.

\subsubsection{Explaining Component Risk Concentration Through Case Studies}\label{sec:rq3:case-illustration}

The representative cases in Section~\ref{sec:rq2} provide concrete evidence 
for interpreting the component-level concentration discussed above.
Rather than serving as isolated examples of mechanism-level failures, 
they help explain why different components occupy different risk-bearing roles in the eBPF architecture.

For \textbf{Runtime}, 
Cases~3--5 provide a direct explanation for why this stage forms the dominant execution exposure surface.
Case~3 shows that inconsistencies between error-recovery logic and 
object release timing can produce dangling references and use-after-free conditions.
Case~4 shows that eBPF-triggered operations may execute under multiple contexts, 
breaking original locking assumptions and exposing concurrency failures.
Case~5 further shows that eBPF can transform subsystem-internal operations into user-triggerable execution paths, 
thereby exposing latent flaws in surrounding kernel subsystems.
Overall, these cases show that Runtime vulnerability concentration is structurally driven 
by concrete execution-time interaction with kernel state, object lifecycles, concurrency, 
and cross-subsystem behavior, rather than by a small number of incidental bugs.

For the \textbf{Verifier}, Case~1 illustrates a different kind of risk.
The failure does not arise from rich runtime interaction, but from incomplete semantic constraints in static safety reasoning.
In this case, the verifier fails to correctly constrain a nullable pointer type, producing a divergence between pointer arithmetic semantics and NULL-branch reasoning, and thereby allowing a program that should have been rejected to be accepted instead.
This shows that the Verifier's architectural risk lies not in frequent execution-time failures, but in the possibility of unsound security judgment at the kernel entry boundary.
When such failures occur, the safety guarantees expected at the enforcement stage are invalidated before execution even begins.

For the \textbf{JIT}, Case~2 highlights yet another distinct risk role.
Here, the problem is not incorrect verifier reasoning, but semantic divergence introduced during architecture-specific translation.
The JIT-generated machine code adds behavior not present in the original bytecode, causing the executed semantics to deviate from those previously approved by the Verifier.
This illustrates that the JIT's architectural significance is not captured by frequency alone: although JIT vulnerabilities are fewer in number, they threaten the semantic-preservation boundary between verified bytecode and machine execution.
Their core risk is therefore the possibility that trusted translation breaks guarantees established earlier in the security pipeline.

Overall, these cases reinforce the component-level interpretation developed in RQ3.
Our findings are consistent with viewing Runtime as the dominant execution exposure surface in practice, 
the Verifier as a critical security-enforcement boundary, 
and the JIT as a semantic-preservation boundary whose failures are less frequent but still structurally important.
The uneven component distribution observed above appears 
to reflect a deeper architectural differentiation across the eBPF execution flow, 
rather than merely a simple imbalance in raw vulnerability counts.

\begin{tcolorbox}[colback=gray!15,colframe=gray!60,boxrule=0.8pt,arc=4pt]
\textbf{TA 4. In the observed dataset, dominant eBPF failure mechanisms are 
unevenly distributed across major components and execution stages. 
The results are consistent with viewing Runtime as the dominant execution exposure surface, 
while the Verifier and JIT contribute smaller but structurally distinct failure classes associated 
with the security-enforcement and semantic-preservation boundaries, respectively. 
Overall, these findings suggest clear architectural differentiation in the observed eBPF risk landscape.}
\end{tcolorbox}

\subsection{RQ4: To what extent do representative existing techniques cover the dominant real-world eBPF
vulnerability patterns?} \label{sec:rq4}

Having established where dominant vulnerabilities concentrate, 
we next examine whether existing representative techniques can effectively cover these failure patterns.
We perform a design-level capability analysis and 
an empirical evaluation using \textit{Syzkaller}~\cite{syzkaller}, augmented by corpus- and semantic-diversity analyses, 
to characterize coverage gaps across the Verifier, JIT, and Runtime.

\subsubsection{Technique Coverage of Dominant Vulnerability Patterns} \label{sec:rq4:design}

\begin{table}[!ht]
    \caption{{\small Mechanism-Level Coverage. \cmark{} = effectively covered; \pmark{} = limited coverage; \xmark{} = not covered.}}
    \label{tab:mechanism-coverage}
    \centering
    \small
    \renewcommand{\arraystretch}{0.8}
    \setlength{\tabcolsep}{4pt}
    \begin{tabularx}{0.9\linewidth}{llXccc}
    \toprule
    \textbf{Category} & \textbf{Scope} & \textbf{Subcategory} & \textbf{Syzkaller} & \textbf{Buzzer} & \textbf{BRF} \\
    \midrule
    
    \multirow{5}{*}{\textbf{A}}
     & \multirow{5}{*}{Verifier}
     & A1: Scalar / range reasoning defects          & \pmark & \pmark & \pmark \\
     & & A2: Pointer / type reasoning defects          & \pmark & \pmark & \pmark \\
     & & A3: State merge / pruning / patching defects   & \pmark & \pmark & \pmark \\
     & & A4: Helper / kfunc behavior modeling defects    & \pmark & \xmark & \pmark \\
     & & A5: Verifier implementation bugs               & \pmark & \xmark & \pmark \\
    
    \midrule
    
    \multirow{4}{*}{\textbf{B}}
     & \multirow{4}{*}{JIT}
     & B1: Instruction selection / encoding errors     & \pmark & \pmark & \pmark \\
     & & B2: Register / stack / calling convention errors & \pmark & \pmark & \pmark \\
     & & B3: Architecture / ABI semantic errors          & \xmark & \xmark & \xmark \\
     & & B4: JIT memory management errors                & \pmark & \xmark & \pmark \\
    
    \midrule
    
    \multirow{6}{*}{\textbf{C}}
     & \multirow{6}{*}{Runtime}
     & C1: Helper implementation defects               & \pmark & \xmark & \cmark \\
     & & C2: Map / storage implementation defects        & \pmark & \xmark & \cmark \\
     & & C3: Kernel object lifecycle / refcount defects   & \pmark & \xmark & \pmark \\
     & & C4: Concurrency / synchronization defects        & \pmark & \xmark & \pmark \\
     & & C5: BPF execution infrastructure defects         & \pmark & \xmark & \pmark \\
     & & C6: Kernel subsystem flaws exposed via eBPF      & \pmark & \xmark & \pmark \\
    
    \midrule
    
    \multirow{3}{*}{\textbf{E}}
     & \multirow{3}{*}{\makecell[l]{Cross-\\component}}
     & E1: Verifier--Runtime mismatch                  & \pmark & \pmark & \pmark \\
     & & E2: JIT--Interpreter mismatch                   & \xmark & \xmark & \xmark \\
     & & E3: Userspace--Kernel ABI mismatch               & \xmark & \xmark & \xmark \\
    
    \bottomrule
    \end{tabularx}
    \end{table}

Table~\ref{tab:mechanism-coverage} summarizes 
the mechanism-level coverage capability of \textit{Syzkaller}~\cite{syzkaller}, Buzzer~\cite{google_buzzer}, 
and BRF~\cite{hung2024brf} across the dominant failure categories identified in RQ2. 
To improve reproducibility, we operationalize the rubric using explicit evidence-based criteria. 
A technique is labeled \textit{effective coverage} (\cmark{}) for a given failure category only if it provides both 
(1) \emph{construction capability}, that is, the ability to generate inputs that can exercise the relevant execution context or mechanism 
(e.g., reaching the required verifier state, JIT path, or runtime interaction), and 
(2) \emph{detection capability}, that is, the ability to expose the failure through observable signals such as crashes, invariant violations, or differential inconsistencies. 
A technique is labeled \textit{limited coverage} (\pmark{}) if it only partially satisfies these conditions. 
This includes cases where the technique can reach the relevant code region but lacks sufficient semantic diversity to reliably exercise the failure mechanism, 
or can construct relevant inputs but lacks effective detection signals for that category. 
A technique is labeled \textit{no clear coverage} (\xmark{}) if neither construction nor detection capability is evident from the design or reported evaluation, 
or if available evidence suggests that the technique is unlikely to exercise the corresponding mechanism in practice. 
When evidence is ambiguous or incomplete, we assign the weaker label. 
All classifications are based on documented design, reported evaluations, and publicly available artifacts when applicable, 
and the rubric was independently applied and cross-checked by multiple authors, with disagreements resolved through discussion.

The overall pattern is uneven: Runtime-related categories receive comparatively stronger support, 
whereas most Verifier-, JIT-, and cross-component categories remain only partially covered or uncovered. 
Notably, JIT architecture/ABI semantic errors (B3) and two cross-component categories (E2 and E3) 
are not covered by any of the three tools. 
Overall, the table indicates that current techniques provide stronger coverage 
for parts of the Runtime space than for many other dominant failure categories.

\vspace{2pt}
\noindent
\textbf{Syzkaller} provides the broadest general-purpose coverage among the three tools. 
As reflected in Table~\ref{tab:mechanism-coverage}, 
it attains at least partial coverage across Verifier, JIT, Runtime, and cross-component categories, 
except for E2 and E3. 
This breadth arises from its syscall-template-based design, which allows it 
to exercise a range of program types, map types, attach types, and common operation sequences, 
thereby reaching many frequently exercised eBPF code paths. 
However, 
its input unit remains a syscall sequence with field-level random mutations, 
without explicit modeling of register abstract states, control-flow structures, or helper-call preconditions. 
This limitation helps explain why its coverage remains partial rather than effective across Verifier and JIT categories, 
and why even within Runtime it does not achieve effective coverage for more demanding mechanism classes.

\vspace{2pt}
\noindent
\textbf{Buzzer} is the only tool among the three that incorporates an active oracle mechanism, 
and its design is focused specifically on the Verifier. 
As shown in Table~\ref{tab:mechanism-coverage}, 
this focus is reflected in partial coverage for the first three Verifier categories (A1--A3), 
but no coverage for Verifier helper/kfunc modeling defects (A4), Verifier implementation bugs (A5), 
any Runtime category, or most JIT and cross-component categories. 
Its program generation remains based on random instruction assembly, 
and its oracle targets a single defect pattern, 
scalar range reasoning errors leading to out-of-bounds pointer arithmetic. 
In addition, 
its support is restricted to the \texttt{SOCKET\_FILTER} program type, \texttt{ARRAY} maps, 
and a small set of helpers. 
These design characteristics are consistent with the coverage profile observed in the table: 
Buzzer provides limited capability within a narrow subset of Verifier-related mechanisms, 
but does not extend effectively to the broader dominant vulnerability space.

\vspace{2pt}
\noindent
\textbf{BRF} shows the strongest Runtime-oriented capability in Table~\ref{tab:mechanism-coverage}. 
It is the only tool marked as effectively covering helper implementation defects (C1) and map/storage implementation defects (C2), 
and it also attains partial coverage across the remaining Runtime categories, 
as well as partial coverage in Verifier, JIT, and E1 cross-component categories. 
This profile is consistent with its design: 
BRF generates structurally complete C programs and compiles them via Clang into valid BPF bytecode, 
substantially increasing the probability that generated programs pass the verifier 
and reach Runtime execution. 
At the same time, 
the table also reflects the limits of this strength. 
BRF remains only partially effective on Verifier-related mechanism categories, 
provides no coverage for the uncovered JIT and cross-component categories, 
and shows no clear support for semantically precise Verifier or JIT failure modes. 
Thus, 
its main advantage lies in stronger Runtime reachability 
rather than broad mechanism-level coverage across the dominant failure space.

The design-level comparison suggests that current techniques 
do not provide uniformly strong coverage across dominant real-world eBPF vulnerability patterns. 
Instead, 
coverage varies substantially by mechanism category: 
some Runtime-related categories receive stronger support, whereas most Verifier-, JIT-, 
and cross-component categories remain only partially covered or uncovered. 
This result motivates the empirical case study of \textit{Syzkaller} in the following subsections, 
where we examine how these capability differences manifest in practice.

\subsubsection{Syzkaller Case Study: Practical Coverage and Discovery Boundaries}
To complement the design-level analysis above, 
we next examine \textit{Syzkaller} as a representative case, 
focusing on its practical coverage behavior and its ability to reach and expose dominant real-world vulnerability patterns.
Specifically, 
we conduct an in-depth 72-hour \textit{Syzkaller} campaign on Linux kernel v5.10.

\begin{figure}[!ht]
\centering
\includegraphics[width=\linewidth]{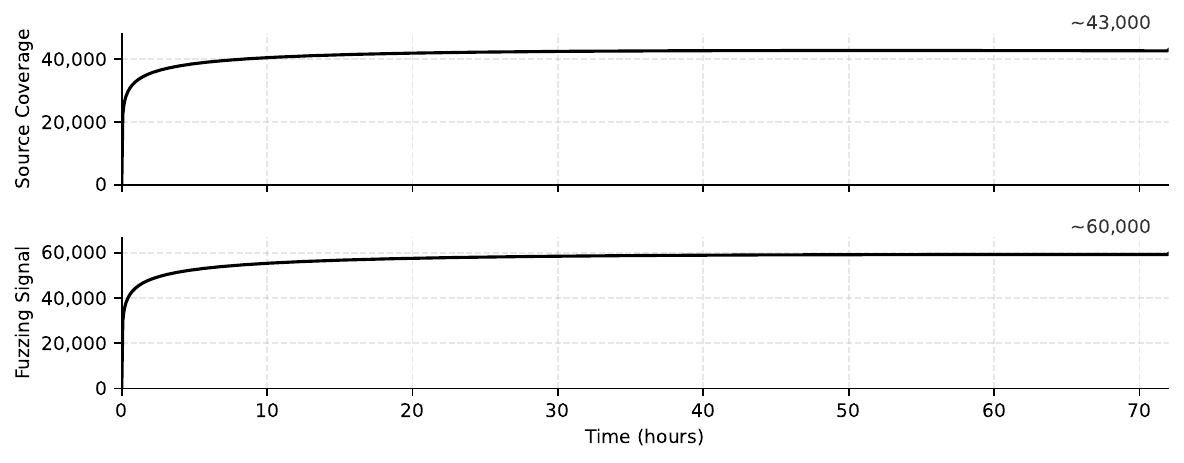}
\caption{Coverage and signal progression of \textit{Syzkaller}. Most growth occurs in the early phase and then quickly stabilizes, suggesting that path exploration saturates rapidly under the current input model.}
\label{fig:coverage_signal}
\end{figure}

\vspace{3pt}
\noindent
\textbf{\blackcircleone{1} Execution Overview and Raw Coverage.}
As shown in Figure~\ref{fig:coverage_signal}, 
both coverage and signal increase rapidly in the early phase and then converge, 
with most growth occurring within the first 10 hours. 
Over the experiment, 
\textit{Syzkaller} executed approximately 140 million test programs but accumulated only 4{,}165 corpus seeds, 
corresponding to about 34{,}000 executions per new-coverage-contributing input. 
This suggests that \textit{Syzkaller} can efficiently exercise frequently triggered paths, 
but that further exploration saturates quickly under the current input model.

\begin{table*}[!ht]
\caption{Per-file KCOV basic-block coverage of core eBPF subsystem components achieved by \textit{Syzkaller}. Coverage is reported as the ratio of executed basic blocks to total instrumented blocks. The results show a structurally uneven coverage distribution across Runtime, Verifier, and JIT components.}
\label{tab:ebpf-file-coverage}
\centering
\small
\renewcommand{\arraystretch}{0.85}
\setlength{\tabcolsep}{4pt}
\begin{tabularx}{\linewidth}{lXp{3.6cm}rrr}
\toprule
\textbf{Component} & \textbf{File} & \textbf{Description} & \textbf{Coverage} & \textbf{Covered} & \textbf{Total} \\
\midrule

\multirow{4}{*}{\textbf{Verifier}}
 & kernel/bpf/verifier.c              & Core verification logic    & 47\% & 1,723 & 3,665 \\
 & kernel/bpf/btf.c                   & BTF type verification      &  1\% &    15 & 1,465 \\
 & kernel/bpf/tnum.c                  & Scalar range tracking      & 74\% &    46 &    61 \\
 & kernel/bpf/disasm.c                & Instruction disassembly    &  5\% &     5 &    82 \\
\cmidrule{1-6}
 & \multicolumn{2}{r}{\textit{Subtotal (Verifier)}} & \textbf{34\%} & \textbf{1,789} & \textbf{5,273} \\

\midrule

\multirow{4}{*}{\textbf{JIT}}
 & arch/x86/net/bpf\_jit\_comp.c      & x86-64 JIT backend         & 53\% & 332 & 626 \\
 & kernel/bpf/core.c                  & Interpreter + JIT dispatch & 15\% & 149 & 992 \\
 & kernel/bpf/trampoline.c            & fentry/fexit trampoline    &  0\% &   0 & 125 \\
 & kernel/bpf/dispatcher.c            & JIT dispatch optimization  &  0\% &   0 &  49 \\
\cmidrule{1-6}
 & \multicolumn{2}{r}{\textit{Subtotal (JIT)}}      & \textbf{27\%} & \textbf{481} & \textbf{1,792} \\

\midrule

\multirow{31}{*}{\textbf{Runtime}}
 & kernel/bpf/syscall.c               & \texttt{bpf()} syscall entry       & 54\% & 732 & 1,356 \\
 & kernel/bpf/hashtab.c               & Hash map                           & 54\% & 348 &   644 \\
 & kernel/bpf/arraymap.c              & Array map                          & 50\% & 226 &   451 \\
 & kernel/bpf/percpu\_freelist.c      & Per-CPU freelist (hashtab)         & 50\% &  28 &    56 \\
 & kernel/bpf/map\_in\_map.c          & Map-in-map support                 & 91\% &  30 &    33 \\
 & kernel/bpf/ringbuf.c               & Ring buffer map                    & 71\% &  60 &    85 \\
 & net/bpf/test\_run.c                & \texttt{BPF\_PROG\_TEST\_RUN}      & 40\% & 128 &   320 \\
 & kernel/bpf/local\_storage.c        & Cgroup local storage               & 28\% &  58 &   208 \\
 & kernel/bpf/cgroup.c                & Cgroup BPF attach/exec             & 15\% & 135 &   899 \\
 & kernel/bpf/inode.c                 & BPF filesystem (bpffs)             & 13\% &  26 &   203 \\
 & kernel/bpf/devmap.c                & Device redirect map                &  9\% &  23 &   258 \\
 & kernel/bpf/reuseport\_array.c      & Reuseport socket map               &  7\% &   9 &   130 \\
 & kernel/bpf/helpers.c               & Generic BPF helpers                &  6\% &  10 &   167 \\
 & kernel/bpf/bpf\_struct\_ops.c      & Struct ops (e.g., TCP CC)          &  6\% &   9 &   155 \\
 & kernel/bpf/queue\_stack\_maps.c    & Queue / stack maps                 &  6\% &   3 &    51 \\
 & net/core/filter.c                  & Network helpers + filters          &  5\% & 171 & 3,427 \\
 & net/core/bpf\_sk\_storage.c        & Socket local storage               &  4\% &  17 &   419 \\
 & net/xdp/xskmap.c                   & XDP socket map                     &  4\% &   3 &    66 \\
 & kernel/bpf/bpf\_local\_storage.c   & Generic local storage infra        &  3\% &   6 &   204 \\
 & kernel/bpf/net\_namespace.c        & Net-namespace BPF ops              &  3\% &   5 &   167 \\
 & kernel/bpf/cpumap.c                & CPU redirect map                   &  2\% &   5 &   250 \\
 & kernel/bpf/stackmap.c              & Stack trace map                    &  2\% &   4 &   208 \\
 & kernel/bpf/lpm\_trie.c             & Longest-prefix-match map           &  2\% &   4 &   205 \\
 & net/core/sock\_map.c               & Socket map / sockhash              &  2\% &  15 &   773 \\
 & kernel/bpf/offload.c               & Hardware offload                   &  1\% &   4 &   434 \\
 & kernel/trace/bpf\_trace.c          & Tracing helpers                    &  1\% &   8 &   802 \\
 & kernel/bpf/bpf\_iter.c             & Iterator framework                 &  0\% &   0 &   178 \\
 & kernel/bpf/bpf\_lru\_list.c        & LRU eviction (hashtab)             &  0\% &   0 &   157 \\
 & kernel/bpf/task\_iter.c            & Task iterator                      &  0\% &   0 &   144 \\
 & kernel/bpf/map\_iter.c             & Map element iterator               &  0\% &   0 &    39 \\
 & kernel/bpf/prog\_iter.c            & Program iterator                   &  0\% &   0 &    14 \\
\cmidrule{1-6}
 & \multicolumn{2}{r}{\textit{Subtotal (Runtime)}}  & \textbf{17\%} & \textbf{2,067} & \textbf{12,503} \\

\midrule
 & \multicolumn{2}{r}{\textbf{Overall}}             & \textbf{22\%} & \textbf{4,337} & \textbf{19,568} \\

\bottomrule
\end{tabularx}
\vspace{2pt}
\end{table*}

At the component level, 
Table~\ref{tab:ebpf-file-coverage} shows an uneven raw-coverage distribution across Runtime, Verifier, and JIT. 
Runtime has the lowest subtotal coverage, reaching only 17\% despite accounting for the largest code volume (12{,}503 PCs). 
Its coverage is also highly polarized: common entry and infrastructure files such as \texttt{syscall.c} (54\%), \texttt{hashtab.c} (54\%), \texttt{arraymap.c} (50\%), 
and \texttt{ringbuf.c} (71\%) are moderately covered, 
whereas many semantically demanding files remain weakly explored, 
including \texttt{filter.c} (5\%), \texttt{helpers.c} (6\%), \texttt{sock\_map.c} (2\%), and \texttt{bpf\_trace.c} (1\%), 
with several iterator- and LRU-related files entirely uncovered. 
Verifier coverage is higher overall at 34\%, with \texttt{verifier.c} reaching 47\% and \texttt{tnum.c} 74\%, 
but this coverage is also selective, since \texttt{btf.c} and \texttt{disasm.c} reach only 1\% and 5\%, respectively. 
JIT reaches 27\% overall: the main backend file \texttt{bpf\_jit\_comp.c} achieves 53\% coverage, 
while \texttt{core.c} reaches only 15\% and both \texttt{trampoline.c} and \texttt{dispatcher.c} remain uncovered. 
These results show that \textit{Syzkaller} reaches visible portions of all three major kernel-side components, 
but that this reach is structurally selective: Runtime coverage is concentrated in frequently exercised infrastructure paths, 
while Verifier and JIT coverage is concentrated in selected core logic rather than broadly distributed across their full semantic surfaces.

\vspace{3pt}
\noindent
\textbf{\blackcircleone{2} Semantic Sparsity Behind Apparent Coverage.}
The raw coverage results above show where \textit{Syzkaller} can reach, 
but not what semantic exploration underlies that reach. 
To examine this, 
we analyze the accumulated corpus from four connected perspectives: 
overall seed distribution and verifier outcomes, 
component-level coverage attribution, 
program-type concentration, 
and semantic-context diversity in covered Verifier and JIT logic.

\vspace{3pt}
\noindent
\textbf{$\Rightarrow$ Overall Seed Distribution and Verifier Outcomes.}
Table~\ref{tab:corpus-overview} summarizes the composition of the 4{,}165 corpus seeds 
accumulated by \textit{Syzkaller}. 
Of these, 
only 642 seeds (15.4\%) contain a \texttt{PROG\_LOAD} operation, 
while the remaining 3{,}523 consist of auxiliary syscalls 
such as map operations, file-descriptor management, and network configuration. 
Among the 642 loading seeds, 
368 (57.3\%) pass the verifier and 274 (42.7\%) are rejected. 
Since JIT compilation and deeper Runtime execution can occur only after successful verification, 
these numbers show that only a small fraction of the effective corpus 
can proceed into later stages of the eBPF execution pipeline.
These results have two implications. 
First, 
verifier passing is already a major gating condition on semantically rich exploration: 
although the corpus contains thousands of coverage-contributing seeds, 
only a limited subset can actually reach JIT compilation and loaded-program execution. 
Second, 
the reported 57.3\% verifier-pass rate should not be interpreted 
as the practical pass rate of \textit{Syzkaller}'s generated inputs overall. 
Because the corpus contains only seeds that contributed new coverage, 
the 642 \texttt{PROG\_LOAD} seeds are already a filtered and enriched subset of the much larger execution population; 
relative to the approximately 140 million executed inputs, 
the effective loading population is extremely small. 
Together, 
these results indicate that \textit{Syzkaller}'s mutation strategy can generate some verifier-passing programs, 
but that successful progression beyond verification remains a substantial bottleneck for deeper exploration of JIT and Runtime behaviors.

\begin{table}[!ht]
\caption{{\small Corpus overview of coverage-contributing seeds generated by \textit{Syzkaller}. The results show that only a small fraction of effective seeds contain \texttt{PROG\_LOAD} and can proceed beyond verification into deeper stages of the eBPF execution pipeline.}}
\label{tab:corpus-overview}
\centering
\small
\renewcommand{\arraystretch}{0.9}
\setlength{\tabcolsep}{5pt}
\begin{tabularx}{0.85\linewidth}{Xr}
\toprule
\textbf{Metric} & \textbf{Value} \\
\midrule
Total executions                       & $\sim$140\,M \\
Corpus seeds (unique signal contributors) & 4{,}165 \\
Exec-to-corpus ratio                   & $\sim$1\,:\,33{,}600 \\
\midrule
Seeds containing BPF operations        & 1{,}653 (39.7\%) \\
\quad Seeds with \texttt{PROG\_LOAD}   & 642 (15.4\%) \\
\quad Seeds with BPF ops but no \texttt{PROG\_LOAD} & 1{,}011 (24.3\%) \\
Seeds without any BPF operation        & 2{,}512 (60.3\%) \\
\midrule
Verifier outcome (642 \texttt{PROG\_LOAD} seeds) & \\
\quad Passed                           & 368 (57.3\%) \\
\quad Rejected                         & 274 (42.7\%) \\
\midrule
Average syscalls per seed               & 3.8 \\
\midrule
Average coverage PCs per seed (all)     & 1{,}743 \\
\quad Verifier-pass group              & 2{,}474 \\
\quad Verifier-reject group            & 1{,}772 \\
\bottomrule
\end{tabularx}
\end{table}

\vspace{3pt}
\noindent
\textbf{$\Rightarrow$ Coverage Attribution Across Components.}
Table~\ref{tab:bpf-coverage-attribution} attributes the covered PCs of core eBPF files to three seed classes: 
verifier-pass seeds, verifier-reject seeds, and seeds without \texttt{PROG\_LOAD}. 
Viewed at the component level, 
the attribution results show clear differentiation across the eBPF execution pipeline.

\begin{table*}[!ht]
\caption{{\small Coverage attribution of core eBPF files by seed class in the \textit{Syzkaller} corpus. Covered PCs are partitioned by verifier-pass seeds, verifier-reject seeds, and seeds without \texttt{PROG\_LOAD}, revealing how different seed classes contribute to coverage across Verifier, JIT, and Runtime components.}}
\label{tab:bpf-coverage-attribution}
\centering
\small
\renewcommand{\arraystretch}{0.85}
\setlength{\tabcolsep}{3pt}
\begin{tabularx}{\linewidth}{lXrrrrrrr}
\toprule
\textbf{Component} & \textbf{File} & \makecell{\textbf{Total}\\\textbf{PCs}} & \makecell{\textbf{Pass}} & \makecell{\textbf{Reject}} & \makecell{\textbf{No-}\\\textbf{Load}} & \makecell{\textbf{Pass}\\\textbf{Only}} & \makecell{\textbf{Reject}\\\textbf{Only}} & \makecell{\textbf{Pass}\\\textbf{$\cap$Reject}} \\
\midrule

\multirow{4}{*}{\textbf{Verifier}}
 & \texttt{kernel/bpf/verifier.c}     & 1{,}690 & 1{,}513 & 1{,}285 & 0   & 405 & 177 & 1{,}108 \\
 & \texttt{kernel/bpf/btf.c}          & 5       & 5       & 5       & 3   & 0   & 0   & 5       \\
 & \texttt{kernel/bpf/tnum.c}         & 45      & 41      & 35      & 0   & 10  & 4   & 31      \\
 & \texttt{kernel/bpf/disasm.c}       & 4       & 4       & 4       & 0   & 0   & 0   & 4       \\

\midrule

\multirow{2}{*}{\textbf{JIT}}
 & \texttt{arch/x86/net/bpf\_jit\_comp.c} & 296 & 296     & 0       & 0   & 296 & 0   & 0       \\
 & \texttt{kernel/bpf/core.c}         & 110     & 110     & 46      & 6   & 64  & 0   & 46      \\

\midrule

\multirow{28}{*}{\textbf{Runtime}}
 & \texttt{kernel/bpf/syscall.c}      & 562     & 394     & 254     & 416 & 171 & 31  & 223     \\
 & \texttt{kernel/bpf/hashtab.c}      & 265     & 103     & 75      & 264 & 29  & 1   & 74      \\
 & \texttt{kernel/bpf/arraymap.c}     & 188     & 108     & 70      & 164 & 46  & 8   & 62      \\
 & \texttt{net/core/filter.c}         & 101     & 91      & 58      & 10  & 43  & 10  & 48      \\
 & \texttt{net/bpf/test\_run.c}       & 81      & 81      & 0       & 0   & 81  & 0   & 0       \\
 & \texttt{kernel/bpf/ringbuf.c}      & 62      & 50      & 13      & 28  & 37  & 0   & 13      \\
 & \texttt{net/xdp/xsk.c}             & 48      & 18      & 9       & 48  & 9   & 0   & 9       \\
 & \texttt{kernel/bpf/cgroup.c}       & 42      & 39      & 35      & 42  & 5   & 1   & 34      \\
 & \texttt{kernel/bpf/local\_storage.c}& 37     & 33      & 0       & 19  & 33  & 0   & 0       \\
 & \texttt{kernel/bpf/percpu\_freelist.c}& 24   & 17      & 17      & 24  & 0   & 0   & 17      \\
 & \texttt{kernel/bpf/inode.c}        & 23      & 13      & 3       & 21  & 10  & 0   & 3       \\
 & \texttt{kernel/bpf/map\_in\_map.c} & 19      & 6       & 6       & 19  & 0   & 0   & 6       \\
 & \texttt{net/core/sock\_map.c}      & 10      & 8       & 0       & 7   & 8   & 0   & 0       \\
 & \texttt{kernel/bpf/devmap.c}       & 9       & 3       & 2       & 9   & 1   & 0   & 2       \\
 & \texttt{kernel/bpf/bpf\_struct\_ops.c}& 6    & 0       & 0       & 6   & 0   & 0   & 0       \\
 & \texttt{net/core/bpf\_sk\_storage.c}& 5      & 5       & 5       & 5   & 0   & 0   & 5       \\
 & \texttt{kernel/bpf/helpers.c}      & 4       & 4       & 2       & 0   & 2   & 0   & 2       \\
 & \texttt{kernel/bpf/offload.c}      & 4       & 3       & 0       & 1   & 3   & 0   & 0       \\
 & \texttt{kernel/bpf/bpf\_local\_storage.c}& 4 & 0       & 0       & 4   & 0   & 0   & 0       \\
 & \texttt{kernel/bpf/net\_namespace.c}& 4      & 4       & 0       & 0   & 4   & 0   & 0       \\
 & \texttt{kernel/bpf/reuseport\_array.c}& 3    & 0       & 0       & 3   & 0   & 0   & 0       \\
 & \texttt{kernel/bpf/cpumap.c}       & 2       & 2       & 0       & 2   & 2   & 0   & 0       \\
 & \texttt{net/xdp/xskmap.c}          & 2       & 0       & 0       & 2   & 0   & 0   & 0       \\
 & \texttt{kernel/trace/bpf\_trace.c} & 2       & 0       & 0       & 2   & 0   & 0   & 0       \\
 & \texttt{kernel/bpf/stackmap.c}     & 2       & 0       & 0       & 2   & 0   & 0   & 0       \\
 & \texttt{net/xdp/xsk\_queue.c}      & 1       & 0       & 0       & 1   & 0   & 0   & 0       \\
 & \texttt{kernel/bpf/lpm\_trie.c}    & 1       & 0       & 0       & 1   & 0   & 0   & 0       \\

\bottomrule
\end{tabularx}
\end{table*}

For the \textbf{Verifier}, 
coverage is contributed by both pass and reject seeds, 
with substantial overlap in \texttt{verifier.c}: 
of the 1{,}285 PCs covered by reject seeds, 1{,}108 are also covered by pass seeds. 
This indicates that many rejected inputs still exercise common verification logic, 
and that a large portion of the observed Verifier coverage comes from shared checking paths rather than semantically distinct exploration.

For the \textbf{JIT}, 
the pattern is more restrictive. 
All 296 covered PCs in \texttt{bpf\_jit\_comp.c} are contributed exclusively by verifier-passing seeds, 
which is consistent with the execution model: JIT compilation can occur only after successful verification. 
Thus, 
JIT coverage depends entirely on the relatively small subset of seeds that survive the Verifier.

For the \textbf{Runtime}, 
the attribution pattern differs again. 
Much of the observed coverage is contributed by seeds without \texttt{PROG\_LOAD}, 
especially in infrastructure-oriented files such as \texttt{syscall.c}, \texttt{hashtab.c}, and \texttt{arraymap.c}. 
For example, 
the No-Load group covers 416 of 562 covered PCs in \texttt{syscall.c}, 264 of 265 in \texttt{hashtab.c}, 
and 164 of 188 in \texttt{arraymap.c}. 
By contrast, 
paths that require loaded-program execution, such as \texttt{test\_run.c}, are covered only by verifier-passing seeds. 
This shows that a substantial portion of the reported Runtime coverage comes from infrastructure-level operations rather than deep semantic execution of BPF programs.

These results clarify the practical meaning of the corpus statistics. 
Verifier coverage is shared across pass and reject seeds, 
JIT coverage depends entirely on the small verifier-passing subset, 
and much of Runtime coverage is driven by no-load infrastructure activity. 
Therefore, 
the raw code coverage reported earlier combines semantically different classes of execution 
and should not be interpreted as uniformly deep traversal of the full eBPF execution pipeline.

\vspace{3pt}
\noindent
\textbf{$\Rightarrow$ Program-Type Distribution of Effective Seeds.}
To further understand why effective loading seeds cover only a limited portion of the eBPF execution pipeline, 
we next examine their program-type composition. 
Table~\ref{tab:by-prog-type} shows that although 13 BPF program types appear among the \texttt{PROG\_LOAD} seeds, 
the distribution is highly concentrated. 
In particular, 
\texttt{CGROUP\_SKB} (390, 60.7\%) and \texttt{SOCKET\_FILTER} (201, 31.3\%) together account for 91\% of all loading seeds, 
while the remaining program types appear only rarely. 
This means that most verifier-passing and verifier-rejected executions are generated from a narrow subset of the available program-type space.

\begin{table}[!ht]
\caption{{\small Distribution of \texttt{PROG\_LOAD} corpus seeds by BPF program type in \textit{Syzkaller}, together with verifier pass/fail outcomes. The table highlights the strong concentration of effective loading seeds in a small number of program types.}}
\label{tab:by-prog-type}
\centering
\small
\renewcommand{\arraystretch}{0.85}
\setlength{\tabcolsep}{3pt}
\begin{tabularx}{\linewidth}{Xlrrrrr}
\toprule
\textbf{Program Type} & \textbf{Description} & \textbf{Seeds} & \makecell{\textbf{With}\\\textbf{PROG\_LOAD}} & \textbf{Pass} & \textbf{Fail} & \makecell{\textbf{Pass Rate}\\\textbf{(PROG\_LOAD)}} \\
\midrule
(none)            & No program loaded        & 3{,}539 & 16  & 0   & 16  & 0.0\%  \\
CGROUP\_SKB       & Cgroup ingress/egress filter & 390 & 390 & 158 & 232 & 40.5\% \\
SOCKET\_FILTER    & Socket packet filter     & 201     & 201 & 201 & 0   & 100.0\%\\
SCHED\_CLS        & Traffic classifier       & 19      & 19  & 1   & 18  & 5.3\%  \\
FLOW\_DISSECTOR   & Flow dissection logic    & 9       & 9   & 9   & 0   & 100.0\%\\
XDP               & eXpress Data Path        & 8       & 8   & 2   & 6   & 25.0\% \\
CGROUP\_SYSCTL    & Cgroup sysctl filter     & 4       & 4   & 0   & 4   & 0.0\%  \\
LWT\_SEG6LOCAL    & Lightweight tunnel SRv6  & 2       & 2   & 2   & 0   & 100.0\%\\
STRUCT\_OPS       & Kernel struct operations & 2       & 2   & 2   & 0   & 100.0\%\\
EXT               & Extension program        & 1       & 1   & 1   & 0   & 100.0\%\\
UNSPEC            & Unspecified type         & 1       & 1   & 1   & 0   & 100.0\%\\
SK\_SKB           & Socket SKB redirect      & 1       & 1   & 0   & 1   & 0.0\%  \\
SK\_REUSEPORT     & Socket reuseport select  & 1       & 1   & 1   & 0   & 100.0\%\\
UNKNOWN(0x1f)     & Invalid/fuzzed type ID   & 1       & 1   & 1   & 0   & 100.0\%\\
\midrule
\textbf{Total}    &                          & \textbf{4{,}165} & \textbf{642} & \textbf{368} & \textbf{274} & \textbf{57.3\%} \\
\bottomrule
\end{tabularx}
\end{table}

The table also shows that this concentration is not uniform in difficulty. 
\texttt{SOCKET\_FILTER} has a 100\% pass rate, while \texttt{CGROUP\_SKB} passes at 40.5\%; 
by contrast, 
types such as \texttt{SCHED\_CLS} (5.3\%), \texttt{XDP} (25.0\%), 
and \texttt{CGROUP\_SYSCTL} (0\%) appear only sparsely and often fail verification. 
Thus, 
the observed corpus is not only narrow in program-type diversity, 
but also biased toward program types that are easier for \textit{Syzkaller}'s 
current template-based generation strategy to construct and pass through verification.
This concentration has direct implications for the semantic meaning of the earlier coverage results. 
The eBPF verifier executes differentiated checking logic for different program types: 
although \texttt{CGROUP\_SKB} and \texttt{SOCKET\_FILTER} can exercise common checking paths, 
many type-specific branches are associated with less represented types 
such as \texttt{XDP}, \texttt{STRUCT\_OPS}, and tracing-related programs. 
When 91\% of the effective loading seeds are concentrated in just two types, 
a large portion of the type-specific semantic checking space remains unlikely to be exercised in practice. 
Therefore, 
the apparent reach observed in the Verifier and JIT should be interpreted 
in light of a strongly skewed program-type distribution rather than as evidence of broad semantic exploration.

\vspace{3pt}
\noindent
\textbf{$\Rightarrow$ Semantic Context Diversity in Verifier and JIT Coverage.}
The program-type concentration above suggests that even covered Verifier and JIT regions 
may be exercised under only a narrow range of semantic contexts. 
To examine this directly, 
we measure, for each covered PC in the Verifier and JIT core files, 
how many distinct \texttt{prog\_type}s reach it. 
To avoid inflating this metric with generic early rejection paths, 
we restrict the analysis to verifier-passing seeds.

\begin{table}[!ht]
\caption{{\small Semantic context diversity of covered PCs in Verifier and JIT core files under verifier-passing \textit{Syzkaller} seeds. Diversity is measured by the number of distinct \texttt{prog\_type}s reaching each covered PC.}}
\label{tab:semantic-diversity}
\centering
\small
\renewcommand{\arraystretch}{0.9}
\setlength{\tabcolsep}{4pt}
\begin{tabularx}{\linewidth}{lXrrrrr}
\toprule
\textbf{Comp.} & \textbf{File} & \makecell{\textbf{Covered}\\\textbf{PCs}} & \makecell{\textbf{1 type}\\\textbf{only}} & \makecell{\textbf{$\leq$2}\\\textbf{types}} & \makecell{\textbf{Max}\\\textbf{(/11)}} & \makecell{\textbf{Average}} \\
\midrule

\multirow{2}{*}{\textbf{Verifier}}
 & \texttt{verifier.c}      & 1{,}513 & 445 (29\%) & 856 (57\%) & 3  & 2.1 \\
 & \texttt{tnum.c}          & 41      & 21 (51\%)  & 33 (80\%)  & 3  & 1.7 \\

\midrule

\textbf{JIT}
 & \texttt{bpf\_jit\_comp.c}& 296     & 75 (25\%)  & 167 (56\%) & 3  & 2.2 \\

\bottomrule
\end{tabularx}
\end{table}

Table~\ref{tab:semantic-diversity} summarizes the results.
In \texttt{verifier.c}, 
445 of the 1{,}513 covered PCs (29\%) are reached by only a single program type, 
and 856 PCs (57\%) are reached by at most two types; 
the average semantic diversity is only 2.1, 
with a maximum of 3 out of 11 observed passing types. 
The same pattern is even more pronounced in \texttt{tnum.c}, 
where 51\% of covered PCs are reached by only one type and 80\% by at most two, with an average of 1.7. 
The JIT backend \texttt{bpf\_jit\_comp.c} shows a similar profile: 
56\% of its covered PCs are reached by at most two types, and the maximum is again only 3. 
These results show that the apparently non-trivial raw coverage reported earlier is supported by only a limited set of semantic contexts.

Combined with the program-type distribution in Table~\ref{tab:by-prog-type}, 
this provides direct evidence that coverage can mask semantic sparsity. 
The 47\% coverage of \texttt{verifier.c} and 
the 53\% coverage of \texttt{bpf\_jit\_comp.c} do not reflect broad exploration of the corresponding semantic spaces; 
instead, 
much of that coverage is driven by common paths exercised repeatedly 
by a small number of dominant program types, especially \texttt{CGROUP\_SKB} and \texttt{SOCKET\_FILTER}. 
Therefore, a covered PC should not be interpreted as evidence that 
the relevant type-specific semantic conditions have been meaningfully explored. 
Rather, 
the observed coverage remains concentrated in a small subset of semantic scenarios, 
leaving many type-specific verification and translation behaviors effectively underexplored.
These results already provide sufficient evidence for the semantic concentration observed in this case study.

\vspace{3pt}
\noindent
\textbf{\blackcircleone{3} Vulnerability Discovery Patterns and Capability Boundaries.}
The preceding analysis shows that \textit{Syzkaller}'s raw coverage 
is supported by only a narrow set of effective seeds and semantic contexts. 
We now compare the defects exposed in the experiment 
against the real-world vulnerability distribution established earlier in our dataset, 
in order to assess how much of the dominant mechanism space \textit{Syzkaller}'s observed findings actually cover.
Table~\ref{tab:syzkaller-crashes} reports the kernel defects detected by \textit{Syzkaller} during the experiment. 
In total, 
only three reproducible crash types were triggered, 
all of them concurrent memory-safety failures in \texttt{kernel/bpf/ringbuf.c}, 
triggered about 290 times in total. 
All observed crashes fall in the Runtime component; 
no Verifier- or JIT-related defects were exposed, despite the visible raw coverage reported earlier 
for both \texttt{verifier.c} (47\%) and \texttt{bpf\_jit\_comp.c} (53\%).

\begin{table}[!ht]
    \caption{{\small Kernel crashes detected by \textit{Syzkaller}. The observed findings are concentrated in a narrow Runtime subregion, with no defects exposed in the Verifier or JIT.}}
    \label{tab:syzkaller-crashes}
    \centering
    \small
    \renewcommand{\arraystretch}{0.9}
    \setlength{\tabcolsep}{5pt}
    \begin{tabular}{lr}
    \toprule
    \textbf{Description} & \textbf{Count} \\
    \midrule
    general protection fault in \_\_bpf\_ringbuf\_reserve & 100 \\
    general protection fault in bpf\_ringbuf\_query       & 100 \\
    general protection fault in corrupted                 & 90  \\
    \bottomrule
    \end{tabular}
\end{table}

The three reproducible crashes share a common root cause: 
one thread invokes \texttt{bpf\_ringbuf\_reserve}, \texttt{bpf\_ringbuf\_} \texttt{output}, 
or \texttt{bpf\_ringbuf\_query} on a ringbuf map through a loaded BPF program, 
while another thread concurrently closes the fd and frees the underlying ringbuf memory, 
causing a use-after-free. 
This corresponds to the C4 concurrency/synchronization category 
in our RQ2 taxonomy and provides concrete evidence that \textit{Syzkaller} can expose 
at least one real Runtime failure pattern from the dominant vulnerability space.
At the same time, 
the observed discovery scope remains limited relative to the ground truth. 
In Linux kernel v5.10, 
our dataset contains 514 real-world eBPF vulnerabilities 
spanning a broad set of dominant weakness classes across Runtime, Verifier, and JIT, 
including memory bounds, race conditions, resource/lifecycle errors, 
initialization defects, locking errors, type-conversion errors, resource consumption, use-after-release, and information exposure. 
Against this ground truth, 
the experiment exposes only three reproducible defects, 
all concentrated in a single Runtime mechanism family. 
Although these crashes are consistent with one dominant class in the dataset, 
most of the observed vulnerability space remains undetected.

This comparison reveals three concrete capability boundaries. 
First, 
discovery is highly concentrated: 
repeated triggering of ringbuf-related crashes indicates rediscovery of a narrow defect pattern 
rather than broad expansion across the dominant mechanism space. 
Second, 
no Verifier- or JIT-related findings are observed, 
even though both components show visible raw coverage; 
relative to the ground-truth distribution, this is a substantial mismatch, 
since Verifier-related mechanisms account for 17.8\% of the sampled population and JIT mechanisms for 8.4\%, 
yet neither is represented in the observed findings. 
Third, 
even within Runtime, where the empirical findings do occur, 
only one mechanism family is hit despite the broader concentration of real-world Runtime vulnerabilities in v5.10. 
These results show that \textit{Syzkaller}'s practical discovery capability 
remains substantially narrower than both its raw coverage numbers and the real-world vulnerability distribution would suggest.

\begin{tcolorbox}[colback=gray!15,colframe=gray!60,boxrule=0.8pt,arc=4pt]
\textbf{TA 5. The rubric-based comparison and the Linux v5.10 Syzkaller case study 
indicate that current eBPF fuzzing support covers the dominant observed vulnerability space only partially. 
Although the examined techniques reach visible portions of the Runtime, Verifier, and JIT at the raw coverage level, 
their effective exploration appears substantially stronger 
for parts of the Runtime space than for several dominant Verifier-, JIT-, and cross-component failure categories. 
In the Syzkaller case study, 
effective exploration remains semantically narrow, 
and observed discoveries are concentrated in a small subset of Runtime failures.}
\end{tcolorbox}

\section{Discussion}\label{sec:discussion}

\subsection{From Empirical Findings to Structural Insights}
Combining the findings across RQ1--RQ4, 
this study provides a unified empirical view of the observed eBPF vulnerability landscape 
and the practical effectiveness of the examined discovery techniques.

\vspace{2pt}
\noindent
\textbf{Structural concentration of vulnerability risk.}
First, 
the observed vulnerability landscape is not uniformly distributed across the system (RQ1, RQ3). 
Instead, it exhibits clear structural concentration: 
Runtime appears as the primary execution exposure surface, 
while the Verifier and JIT represent lower-frequency but semantically critical boundaries 
where failures can invalidate broader safety guarantees.

\vspace{2pt}
\noindent
\textbf{Vulnerability mechanisms are inherently semantic.}
Second, 
the mechanism-level analysis suggests that many dominant eBPF failures are inherently semantic (RQ2). 
Many defects arise only under specific combinations of program types, helper usage, execution contexts, 
and cross-component interactions. 
This suggests that the observed vulnerability landscape depends importantly on semantic conditions 
rather than purely structural code patterns.

\vspace{2pt}
\noindent
\textbf{Effective discovery remains narrower than raw coverage.}
Third, 
the examined techniques provide only partial and uneven coverage of this observed landscape (RQ4). 
Although raw coverage reaches visible portions of the Runtime, Verifier, and JIT, 
effective exploration remains semantically narrow, 
and observed discoveries in the \textit{Syzkaller} case study are concentrated in a small subset of Runtime failures. 
Relative to the version-aligned vulnerability distribution reconstructed for Linux v5.10, 
many dominant Verifier-, JIT-, and Runtime-related mechanisms remain weakly explored.

\subsection{Structural Blind Spots Revealed by the Study}

Synthesizing the findings across RQ1--RQ4, 
our results suggest four structural blind spots in current eBPF vulnerability discovery techniques. 
Rather than being abstract design concerns, 
these blind spots are reflected in the empirical patterns observed in our dataset analysis 
and in the \textit{Syzkaller} case study.

\vspace{2pt}
\noindent
\textbf{Limited Effective Entry into the Full eBPF Execution Pipeline.}
RQ4 shows that only a small fraction of effective seeds actually enter the later stages of eBPF execution: 
among 4{,}165 coverage-contributing seeds, 
only 642 contain \texttt{PROG\_LOAD}, and only 368 of those pass the verifier. 
Since JIT compilation and deeper Runtime execution require successful verification, 
only a limited subset of generated inputs can proceed into the stages 
where many semantically rich failure mechanisms arise. 
Viewed together with RQ3, 
which showed that dominant risks are distributed across Runtime, Verifier, and JIT, 
this suggests a first blind spot: current techniques have limited effective entry into the full eBPF execution pipeline, 
especially beyond shallow infrastructure-level execution.

\vspace{2pt}
\noindent
\textbf{Coverage is Dominated by Structurally Easier Execution Classes.}
RQ4 further shows that the observed code coverage is contributed by semantically different classes of seeds. 
Verifier coverage is shared heavily between pass and reject seeds, 
JIT coverage depends entirely on verifier-passing seeds, 
and much of Runtime coverage is contributed by no-load seeds through infrastructure-level operations 
such as map creation and file-descriptor management. 
Viewed together with RQ3, 
which identified Runtime as the primary execution exposure surface, 
this suggests that visible coverage in the dominant component is often driven by structurally easier execution classes 
rather than by deep end-to-end BPF program behavior. 
The resulting blind spot is that raw coverage can overstate 
how much of the vulnerability-relevant execution space is actually being explored.

\vspace{2pt}
\noindent
\textbf{Severe Semantic Concentration of Effective Inputs.}
RQ2 showed that dominant eBPF vulnerabilities are tied to specific semantic mechanisms 
rather than to generic structural code patterns alone. 
RQ4 then shows that the effective input space remains highly concentrated: 
among \texttt{PROG\_LOAD} seeds, \texttt{CGROUP\_SKB} and \texttt{SOCKET\_FILTER} account for 91\% of the population, 
and most covered PCs in the Verifier and JIT are reached by only one or two program types. 
This means that even when current techniques do enter verifier-driven execution, 
they do so under only a narrow subset of semantic contexts. 
The corresponding blind spot is therefore not merely low diversity in a generic sense, 
but a concrete concentration of exploration in a small number of semantic scenarios, 
leaving many type-specific checking and translation behaviors underexplored.

\vspace{2pt}
\noindent
\textbf{Practical Discovery Remains Narrow Relative to the Observed Vulnerability Landscape.}
Finally, 
the combined findings of RQ1 and RQ4 reveal a direct gap between observed vulnerability prevalence 
and practical discovery capability. 
RQ1 established that the Linux kernel v5.10 slice of our reconstructed dataset contains 514 eBPF vulnerability instances 
spanning diverse mechanisms across Runtime, Verifier, and JIT. 
However, RQ4 shows that the \textit{Syzkaller} case study exposes only a very small number of defects, 
all concentrated in a single Runtime concurrency-related family, 
with no observed Verifier- or JIT-related findings despite visible raw coverage in both components. 
In light of RQ2 and RQ3, 
which showed that dominant failure mechanisms and architectural risk are distributed more broadly, 
this comparison suggests a final blind spot: current techniques can accumulate visible code coverage 
without achieving correspondingly broad vulnerability discovery across the observed mechanism space.

\subsection{Implications for Future eBPF Vulnerability Discovery}

Although our results are empirical rather than prescriptive, 
they suggest several directions that may be useful for improving future eBPF vulnerability-discovery techniques.

\vspace{2pt}
\noindent
\textbf{Semantic coverage as a complementary objective.}
The observed concentration of effective seeds and semantic contexts suggests that future input generation 
may benefit from incorporating semantic coverage objectives alongside syntactic mutation. 
In practice, 
this could include more systematic exploration of dimensions such as program type, helper usage, map configuration, 
attach context, and execution environment, 
so that generated inputs are more likely to exercise semantically distinct behaviors 
across the Verifier, JIT, and Runtime.

\vspace{2pt}
\noindent
\textbf{Deeper and more specialized exploration.}
The gap between visible raw coverage and narrow practical discovery suggests that future techniques 
may benefit from stronger support for deeper and more specialized exploration. 
Many observed eBPF vulnerabilities appear to depend on multi-step interactions involving helper calls, 
resource lifecycles, concurrency conditions, or subsystem-specific execution states. 
This suggests that techniques capable of constructing richer interaction sequences 
or selectively focusing on higher-risk semantic regions 
may offer broader discovery potential than uniformly coverage-driven exploration alone.

\vspace{2pt}
\noindent
\textbf{Beyond crash-based feedback.}
The absence of Verifier- and JIT-related findings despite visible code reach suggests 
that crash-based feedback alone may be insufficient for some important classes of eBPF defects. 
One promising direction is to augment existing fuzzing workflows with additional semantic detection mechanisms, 
such as invariant checking, differential validation, 
or cross-stage consistency checks between verifier assumptions, JIT output, and runtime behavior. 
Our results do not directly evaluate such mechanisms, 
but they suggest that some form of semantic feedback may be important 
for improving visibility into non-crash defect spaces.

\vspace{2pt}
\noindent
\textbf{Component-aware and pipeline-aware testing strategies.}
The differing failure characteristics across Runtime, Verifier, and JIT suggest 
that future techniques may benefit from more component-aware or pipeline-aware testing strategies. 
Rather than treating all parts of the eBPF stack as equivalent fuzzing targets, 
it may be useful to adapt input generation, guidance, and checking logic 
to the distinct semantic roles of different execution stages, 
while also paying more explicit attention to cross-component interactions.
\section{Threats to Validity} \label{sec:threats}

Our study has several limitations that should be considered when interpreting the results. 
For each, we briefly note both the potential threat and the steps taken to reduce its impact.

\noindent
\textbf{Dataset construction and labeling.}
Our empirical analyses depend on the completeness and consistency of the vulnerability dataset. 
Public disclosures may be incomplete, some vulnerabilities may be underreported, 
and root-cause categorization can be ambiguous for cases spanning multiple mechanisms or components. 
To mitigate this threat, 
we used a structured collection and analysis process, 
applied a consistent taxonomy across RQ1--RQ3, 
and interpreted component- and mechanism-level labels using uniform criteria. 
However, some borderline cases may still admit alternative categorizations, 
and the dataset should be interpreted as a best-effort empirical view rather than a complete ground truth.
In addition, because kernel fixing commits account for most retained instances, 
the dataset may reflect biases from maintainer practices, patch structure, and commit-message conventions.
Although cross-source linking reduces direct duplication, 
some residual fragmentation may remain when multiple commits or incompletely linked records correspond to a single issue.
Finally, although we performed independent labeling on a subset and adjudicated disagreements, 
the mechanism taxonomy should be interpreted as a structured analytical abstraction with bounded validation 
rather than a fully externally validated ontology.
The dataset represents a best-effort empirical reconstruction of observed eBPF vulnerability instances, 
with limitations in source balance, record consolidation, and classification granularity.

\noindent
\textbf{Version alignment and scope of ground truth.}
Our comparison between observed findings and real-world vulnerabilities is 
grounded in Linux kernel v5.10. 
This provides a concrete and version-aligned basis for RQ4, 
but vulnerability distributions may differ across kernel versions due to code evolution, 
subsystem refactoring, backports, and deployment differences. 
We mitigate this threat by explicitly aligning both the dataset comparison and the \textit{Syzkaller} case study 
to the same target version. 
Nevertheless, 
the quantitative comparison should be interpreted as version-specific 
rather than representative of all eBPF-enabled kernels.

\noindent
\textbf{Generality of the \textit{Syzkaller} case study.}
The empirical part of RQ4 centers on \textit{Syzkaller} 
as a representative general-purpose dynamic technique. 
While this choice is motivated by its practical importance and widespread use, 
a single case study cannot fully represent the behavior of all existing or future eBPF vulnerability-discovery techniques. 
To mitigate this threat, 
we complement the case study with a design-level comparison of Buzzer and BRF, 
so that RQ4 is not based solely on one empirical system. 
Still, 
the detailed observations on seed effectiveness, semantic sparsity, and observed discovery patterns 
should be interpreted primarily as evidence about \textit{Syzkaller}, 
and more cautiously as indicative of broader challenges in eBPF vulnerability discovery.

\noindent
\textbf{Coverage and measurement limitations.}
Our study relies on KCOV-based basic-block coverage, 
corpus replay, seed attribution, program-type distribution, 
and semantic-context analysis to characterize exploration behavior. 
These metrics are informative, 
but none provides a complete measure of semantic reachability or defect-triggering capability. 
In particular, 
a covered program counter does not necessarily imply that 
the corresponding semantic scenario has been meaningfully explored. 
We mitigate this threat by combining multiple complementary analyses 
rather than relying on raw coverage alone. 
However, 
all coverage-oriented measurements remain imperfect proxies for vulnerability-discovery capability in this study.

\noindent
\textbf{Experimental environment and fuzzing budget.}
The results of fuzzing campaigns depend on kernel configuration, 
architecture, runtime environment, randomness, scheduler behavior, and campaign duration. 
Different setups or longer campaigns may lead to different outcomes. 
To reduce this threat, 
we use a fixed and explicitly defined experimental setup and 
evaluate \textit{Syzkaller} under a sustained 72-hour campaign, 
providing a controlled basis for comparison. 
Even so, 
our conclusions should be interpreted as characterizing practical capability under the evaluated setup, 
not as upper bounds on what any technique could achieve under all configurations.
\section{Conclusion}

This paper presented a systematic empirical study of observed eBPF vulnerabilities. 
Rather than examining individual Verifier, JIT, or Runtime mechanisms in isolation, 
we analyzed a reconstructed empirical view of the eBPF vulnerability landscape grounded in Linux kernel fixing commits, syzbot reports, and public vulnerability records. 
Our results suggest that the observed landscape is not uniformly distributed across the system, 
but instead exhibits clear structural concentration across a limited number of dominant weakness categories and architectural regions.

Building on this structural view, 
we found that the dominant observed portion of the landscape is associated with recurring system-level failure mechanisms rather than a broad collection of unrelated one-off defects. 
These mechanisms are themselves non-uniformly distributed across the eBPF architecture, 
with Runtime-related failures occupying the largest observed portion, 
alongside smaller but structurally distinct Verifier-, JIT-, and cross-component issues. 
When compared against this empirical structure, 
representative current discovery techniques provide only partial coverage in practice: 
although they reach visible portions of the Verifier, JIT, and Runtime at the raw coverage level, 
their effective exploration remains semantically narrow, 
and observed discoveries are concentrated in a limited subset of Runtime-related failures.

Overall, 
these findings suggest that the observed eBPF vulnerability landscape is better understood as a structured system-level problem than as a flat collection of evenly distributed implementation-level errors. 
They also indicate that raw reachability or isolated bug-finding success can provide an incomplete view of practical discovery effectiveness. 
By connecting structural concentration, failure mechanisms, architectural distribution, and discovery gaps within a unified empirical framework, 
this study helps clarify where current eBPF security analysis is effective, where it remains limited, 
and which parts of the observed landscape may require stronger support from future analysis and testing techniques.

\bibliographystyle{ACM-Reference-Format}
\bibliography{reference}

@STRING{ASE     = "Proceedings of IEEE/ACM International Conference on Automated Software Engineering"}

@STRING{FSE     = "Proceedings of ACM International Symposium on the Foundations of Software Engineering"}

@article{hung2024brf,
  title={Brf: Fuzzing the ebpf runtime},
  author={Hung, Hsin-Wei and Amiri Sani, Ardalan},
  journal={Proceedings of the ACM on Software Engineering},
  volume={1},
  number={FSE},
  pages={1152--1171},
  year={2024},
  publisher={ACM New York, NY, USA}
}

@MISC{syzkaller,
  author = {{Google}},
  title = {{syzkaller: Kernel Fuzzer}},
  howpublished = {\url{https://github.com/google/syzkaller}},
  year = {2024}
}

@MISC{nvd,
  author = {{National Institute of Standards and Technology}},
  title = {{National Vulnerability Database}},
  howpublished = {\url{https://nvd.nist.gov/}},
  year = {2024}
}

@MISC{cwe1000,
  author = {{MITRE}},
  title = {{CWE-1000: Research Concepts View}},
  howpublished = {\url{https://cwe.mitre.org/data/definitions/1000.html}},
  year = {2024}
}

@MISC{linuxkernel,
  author = {{Linux Kernel Community}},
  title = {{Linux Kernel Source Code}},
  howpublished = {\url{https://github.com/torvalds/linux}},
  year = {2024}
}

@MISC{syzbot,
  author = {{Google}},
  title = {{syzbot: Continuous Kernel Bug Finding System}},
  howpublished = {\url{https://syzkaller.appspot.com/}},
  year = {2024}
}

@article{cohen1960kappa,
  author = {Jacob Cohen},
  title = {A Coefficient of Agreement for Nominal Scales},
  journal = {Educational and Psychological Measurement},
  year = {1960},
  volume = {20},
  number = {1},
  pages = {37--46}
}

@article{fleiss1971kappa,
  author = {Joseph L. Fleiss},
  title = {Measuring Nominal Scale Agreement Among Many Raters},
  journal = {Psychological Bulletin},
  year = {1971},
  volume = {76},
  number = {5},
  pages = {378--382}
}

@misc{cwe_mapping_guidance,
  author = {{MITRE}},
  title = {{CVE to CWE Root Cause Mapping Guidance}},
  howpublished = {\url{https://cwe.mitre.org/documents/cwe_usage/guidance.html}},
  year = {2024}
}

@book{neuendorf2002content,
  author = {Kimberly A. Neuendorf},
  title = {The Content Analysis Guidebook},
  publisher = {SAGE Publications},
  year = {2002}
}

@book{Cochran1977,
  author    = {William G. Cochran},
  title     = {Sampling Techniques},
  edition   = {3},
  year      = {1977},
  publisher = {Wiley},
  address   = {New York},
  isbn      = {047116240X},
  url       = {https://www.wiley.com/en-us/Sampling+Techniques%2C+3rd+Edition-p-9780471162407}
}

@book{lohr2021sampling,
  title={Sampling: design and analysis},
  author={Lohr, Sharon L},
  year={2021},
  publisher={Chapman and Hall/CRC}
}

@inproceedings{gershuni2019simple,
  title={Simple and precise static analysis of untrusted linux kernel extensions},
  author={Gershuni, Elazar and Amit, Nadav and Gurfinkel, Arie and Narodytska, Nina and Navas, Jorge A and Rinetzky, Noam and Ryzhyk, Leonid and Sagiv, Mooly},
  booktitle={Proceedings of the 40th ACM SIGPLAN Conference on Programming Language Design and Implementation},
  pages={1069--1084},
  year={2019}
}

@inproceedings{sun2024validating,
  title={Validating the $\{$eBPF$\}$ verifier via state embedding},
  author={Sun, Hao and Su, Zhendong},
  booktitle={18th USENIX Symposium on Operating Systems Design and Implementation (OSDI 24)},
  pages={615--628},
  year={2024}
}

@inproceedings{nelson2020specification,
  title={Specification and verification in the field: Applying formal methods to $\{$BPF$\}$ just-in-time compilers in the linux kernel},
  author={Nelson, Luke and Van Geffen, Jacob and Torlak, Emina and Wang, Xi},
  booktitle={14th USENIX Symposium on Operating Systems Design and Implementation (OSDI 20)},
  pages={41--61},
  year={2020}
}

@article{bhat2022formal,
  title={Formal verification of the linux kernel eBPF verifier range analysis},
  author={Bhat, Sanjit and Shacham, Hovav},
  journal={Semantic Scholar (2022). https://api. semanticscholar. org/CorpusID},
  volume={252564197},
  year={2022}
}

@inproceedings{jia2025rex,
  title={Rex: Closing the language-verifier gap with safe and usable kernel extensions},
  author={Jia, Jinghao and Qin, Ruowen and Craun, Milo and Lukiyanov, Egor and Bansal, Ayush and Phan, Minh and Le, Michael V and Franke, Hubertus and Jamjoom, Hani and Xu, Tianyin and others},
  booktitle={2025 USENIX Annual Technical Conference (USENIX ATC 25)},
  pages={325--342},
  year={2025}
}

@inproceedings{peng2024toss,
  title={Toss a fault to bpfchecker: Revealing implementation flaws for ebpf runtimes with differential fuzzing},
  author={Peng, Chaoyuan and Jiang, Muhui and Wu, Lei and Zhou, Yajin},
  booktitle={Proceedings of the 2024 on ACM SIGSAC Conference on Computer and Communications Security},
  pages={3928--3942},
  year={2024}
}

@inproceedings{huang2025sok,
  title={Sok: Challenges and paths toward memory safety for ebpf},
  author={Huang, Kaiming and Payer, Mathias and Qian, Zhiyun and Sampson, Jack and Tan, Gang and Jaeger, Trent},
  booktitle={2025 IEEE Symposium on Security and Privacy (SP)},
  pages={848--866},
  year={2025},
  organization={IEEE}
}

@misc{ebpfdocs,
  title        = {eBPF Documentation},
  author       = {{eBPF Foundation}},
  year         = {2024},
  howpublished = {\url{https://docs.ebpf.io}},
  note         = {Accessed: 2026-04-09}
}

@software{google_buzzer,
  author       = {{Google}},
  title        = {Buzzer: An eBPF Fuzzer Toolchain},
  year         = {2023},
  url          = {https://github.com/google/buzzer},
  note         = {Software repository. Accessed: 2026-04-10}
}

@inproceedings{li2017securitypatches,
  author    = {Li, Frank and Paxson, Vern},
  title     = {A Large-Scale Empirical Study of Security Patches},
  booktitle = {Proceedings of the 2017 ACM SIGSAC Conference on Computer and Communications Security},
  year      = {2017}
}

@inproceedings{8990271,
  author    = {Ruohonen, Jukka and Rindell, Kalle},
  title     = {Empirical Notes on the Interaction Between Continuous Kernel Fuzzing and Development},
  booktitle = {2019 IEEE International Symposium on Software Reliability Engineering Workshops (ISSREW)},
  pages     = {276--281},
  year      = {2019}
}

@INPROCEEDINGS{11352476,
  author={Bursey, Joseph and Sani, Ardalan Amiri and Qian, Zhiyun},
  booktitle={2025 28th International Symposium on Research in Attacks, Intrusions and Defenses (RAID)}, 
  title={SyzRetrospector: A Large-Scale Retrospective Study of Syzbot}, 
  year={2025},
  volume={},
  number={},
  pages={92-105}}

@inproceedings {277242,
author = {Xiaochen Zou and Guoren Li and Weiteng Chen and Hang Zhang and Zhiyun Qian},
title = {{SyzScope}: Revealing {High-Risk} Security Impacts of {Fuzzer-Exposed} Bugs in Linux kernel},
booktitle = {31st USENIX Security Symposium (USENIX Security 22)},
year = {2022},
isbn = {978-1-939133-31-1},
address = {Boston, MA},
pages = {3201--3217},
url = {https://www.usenix.org/conference/usenixsecurity22/presentation/zou},
publisher = {USENIX Association},
month = aug
}

@article{liu2025disclosure,
  author  = {Liu, Shuhan and Zhou, Jiayuan and Hu, Xing and Cogo, Filipe Roseiro and Xia, Xin and Yang, Xiaohu},
  title   = {An Empirical Study on Vulnerability Disclosure Management of Open Source Software Systems},
  journal = {ACM Transactions on Software Engineering and Methodology},
  volume  = {34},
  number  = {7},
  year    = {2025}
}

@inproceedings{liu2020vulnerabilitydistribution,
  author    = {Liu, Bingchang and Meng, Guozhu and Zou, Wei and Gong, Qi and Li, Feng and Lin, Min and Sun, Dandan and Huo, Wei and Zhang, Chao},
  title     = {A large-scale empirical study on vulnerability distribution within projects and the lessons learned},
  booktitle = {Proceedings of the ACM/IEEE 42nd International Conference on Software Engineering},
  year      = {2020}
}

@inproceedings{akhoundali2024morefixes,
  author    = {Akhoundali, Jafar and Nouri, Sajad Rahim and Rietveld, Kristian and Gadyatskaya, Olga},
  title     = {MoreFixes: A Large-Scale Dataset of CVE Fix Commits Mined through Enhanced Repository Discovery},
  booktitle = {Proceedings of the 20th International Conference on Predictive Models and Data Analytics in Software Engineering},
  pages     = {42--51},
  year      = {2024}
}

@inproceedings {281444,
author = {Bodong Zhao and Zheming Li and Shisong Qin and Zheyu Ma and Ming Yuan and Wenyu Zhu and Zhihong Tian and Chao Zhang},
title = {{StateFuzz}: System {Call-Based} {State-Aware} Linux Driver Fuzzing},
booktitle = {31st USENIX Security Symposium (USENIX Security 22)},
year = {2022},
isbn = {978-1-939133-31-1},
address = {Boston, MA},
pages = {3273--3289},
url = {https://www.usenix.org/conference/usenixsecurity22/presentation/zhao-bodong},
publisher = {USENIX Association},
month = aug
}

@inproceedings {291291,
author = {Marius Fleischer and Dipanjan Das and Priyanka Bose and Weiheng Bai and Kangjie Lu and Mathias Payer and Christopher Kruegel and Giovanni Vigna},
title = {{ACTOR}: {Action-Guided} Kernel Fuzzing},
booktitle = {32nd USENIX Security Symposium (USENIX Security 23)},
year = {2023},
isbn = {978-1-939133-37-3},
address = {Anaheim, CA},
pages = {5003--5020},
url = {https://www.usenix.org/conference/usenixsecurity23/presentation/fleischer},
publisher = {USENIX Association},
month = aug
}

@inproceedings {280702,
author = {Hao Sun and Yuheng Shen and Jianzhong Liu and Yiru Xu and Yu Jiang},
title = {{KSG}: Augmenting Kernel Fuzzing with System Call Specification Generation},
booktitle = {2022 USENIX Annual Technical Conference (USENIX ATC 22)},
year = {2022},
isbn = {978-1-939133-29-20},
address = {Carlsbad, CA},
pages = {351--366},
url = {https://www.usenix.org/conference/atc22/presentation/sun},
publisher = {USENIX Association},
month = jul
}

@inproceedings {291011,
author = {Philipp G{\"o}rz and Bj{\"o}rn Mathis and Keno Hassler and Emre G{\"u}ler and Thorsten Holz and Andreas Zeller and Rahul Gopinath},
title = {Systematic Assessment of Fuzzers using Mutation Analysis},
booktitle = {32nd USENIX Security Symposium (USENIX Security 23)},
year = {2023},
isbn = {978-1-939133-37-3},
address = {Anaheim, CA},
pages = {4535--4552},
url = {https://www.usenix.org/conference/usenixsecurity23/presentation/gorz},
publisher = {USENIX Association},
month = aug
}

@inproceedings{shi2019enterprisefuzzing,
  author    = {Shi, Heyuan and Wang, Runzhe and Fu, Ying and Wang, Mingzhe and Shi, Xiaohai and Jiao, Xun and Song, Houbing and Jiang, Yu and Sun, Jiaguang},
  title     = {Industry practice of coverage-guided enterprise Linux kernel fuzzing},
  booktitle = {Proceedings of the 2019 27th ACM Joint Meeting on European Software Engineering Conference and Symposium on the Foundations of Software Engineering},
  year      = {2019}
}

@inproceedings{sun2024ebpfverifier,
  author    = {Sun, Hao and Xu, Yiru and Liu, Jianzhong and Shen, Yuheng and Guan, Nan and Jiang, Yu},
  title     = {Finding Correctness Bugs in eBPF Verifier with Structured and Sanitized Program},
  booktitle = {Proceedings of the 39th ACM/SIGAPP Symposium on Applied Computing},
  year      = {2024}
}

@inproceedings{mohamed2023ebpfsecurity,
  author    = {Mohamed, Mohamed Husain Noor and Wang, Xiaoguang and Ravindran, Binoy},
  title     = {Understanding the Security of Linux eBPF Subsystem},
  booktitle = {Proceedings of the 14th ACM SIGOPS Asia-Pacific Workshop on Systems},
  year      = {2023}
}

@inproceedings{hao2022dependencychallenge,
  author    = {Hao, Yu and Zhang, Hang and Li, Guoren and Du, Xingyun and Qian, Zhiyun and Sani, Ardalan Amiri},
  title     = {Demystifying the Dependency Challenge in Kernel Fuzzing},
  booktitle = {Proceedings of the 31st ACM SIGSOFT International Symposium on Software Testing and Analysis},
  year      = {2022}
}

@inproceedings {217573,
author = {Shankara Pailoor and Andrew Aday and Suman Jana},
title = {{MoonShine}: Optimizing {OS} Fuzzer Seed Selection with Trace Distillation},
booktitle = {27th USENIX Security Symposium (USENIX Security 18)},
year = {2018},
isbn = {978-1-939133-04-5},
address = {Baltimore, MD},
pages = {729--743},
url = {https://www.usenix.org/conference/usenixsecurity18/presentation/pailoor},
publisher = {USENIX Association},
month = aug
}

@inproceedings{bhandari2021cvefixes,
  author    = {Bhandari, Guru and Naseer, Amara and Moonen, Leon},
  title     = {CVEfixes: Automated Collection of Vulnerabilities and Their Fixes from Open-Source Software},
  booktitle = {Proceedings of the 17th International Conference on Predictive Models and Data Analytics in Software Engineering},
  year      = {2021}
}

@INPROCEEDINGS{spinner,
  author={Govindasamy, Priya and Bursey, Joseph and Hung, Hsin-Wei and Sani, Ardalan Amiri},
  booktitle={2025 40th IEEE/ACM International Conference on Automated Software Engineering (ASE)}, 
  title={Spinner: Detecting Locking Violations in the eBPF Runtime}, 
  year={2025},
  volume={},
  number={},
  pages={2171-2183}}

@inproceedings{hao2025syzspec,
  author    = {Hao, Yu and Pu, Juefei and Li, Xingyu and Qian, Zhiyun and Sani, Ardalan Amiri},
  title     = {SyzSpec: Specification Generation for Linux Kernel Fuzzing via Under-Constrained Symbolic Execution},
  booktitle = {Proceedings of the 2025 ACM SIGSAC Conference on Computer and Communications Security},
  year      = {2025}
}

@inproceedings{lyu2025ebpfmisbehavior,
  author    = {Lyu, Tao and Dwivedi, Kumar Kartikeya and Bourgeat, Thomas and Payer, Mathias and Xu, Meng and Kashyap, Sanidhya},
  title     = {eBPF Misbehavior Detection: Fuzzing with a Specification-Based Oracle},
  booktitle = {Proceedings of the 2025 ACM SIGSAC Conference on Computer and Communications Security},
  year      = {2025}
}

@inproceedings{zhong2025depsurf,
  author    = {Zhong, Shawn Wanxiang and Liu, Jing and Arpaci-Dusseau, Andrea and Arpaci-Dusseau, Remzi},
  title     = {Revealing the Unstable Foundations of eBPF-Based Kernel Extensions},
  booktitle = {Proceedings of the Twentieth European Conference on Computer Systems},
  pages     = {21--41},
  year      = {2025}
}

@inproceedings{pauley2023cve,
  author = {Pauley, Eric and Barford, Paul and McDaniel, Patrick},
  title = {The CVE Wayback Machine: Measuring Coordinated Disclosure from Exploits against Two Years of Zero-Days},
  year = {2023}
}

@inproceedings{wang2024reposvul,
  author = {Wang, Xinchen and Hu, Ruida and Gao, Cuiyun and Wen, Xin-Cheng and Chen, Yujia and Liao, Qing},
  title = {ReposVul: A Repository-Level High-Quality Vulnerability Dataset},
  booktitle = {Proceedings of the 2024 IEEE/ACM 46th International Conference on Software Engineering: Companion Proceedings},
  year = {2024},
  pages = {472--483},
  doi = {10.1145/3639478.3647634}
}

@inproceedings {299527,
author = {Zheng Zhang and Yu Hao and Weiteng Chen and Xiaochen Zou and Xingyu Li and Haonan Li and Yizhuo Zhai and Billy Lau},
title = {{SymBisect}: Accurate Bisection for {Fuzzer-Exposed} Vulnerabilities},
booktitle = {33rd USENIX Security Symposium (USENIX Security 24)},
year = {2024},
isbn = {978-1-939133-44-1},
address = {Philadelphia, PA},
pages = {2493--2510},
url = {https://www.usenix.org/conference/usenixsecurity24/presentation/zhang-zheng},
publisher = {USENIX Association},
month = aug
}


\end{document}